\documentclass[10pt,journal,compsoc]{IEEEtran}
\ifCLASSOPTIONcompsoc
  \usepackage[nocompress]{cite}
\else
  \usepackage{cite}
\fi
\ifCLASSINFOpdf
  \usepackage[pdftex]{graphicx}
\else
\fi

\usepackage{ifthen}
\newboolean{VersionForJournal}
\setboolean{VersionForJournal}{false}
\newboolean{IncludeAppendix}
\setboolean{IncludeAppendix}{true}

\newcommand{\variable}[1]{{\bf #1}}   %
\newcommand{\condition}[1]{#1}   %
\newcommand{\MainCondition}{\variable{MainCondition}}
\newcommand{\VRTwo}{\condition{VR2D}} %
\newcommand{\VRTwoHilite}{\condition{VR2DHilite}} %
\newcommand{\VRThree}{\condition{VR3D}} %
\newcommand{\VRThreeHilite}{\condition{VR3DHilite}} %
\newcommand{\AR}{\condition{AR3D}} %
\newcommand{\ARTouch}{\condition{AR3DTouch}} %

\newcommand{\ARHilite}{\condition{AR3DHilite}} %

\newcommand{\PathLength}{\variable{PathLength}}

\begin{document}

\title{Path Tracing in 2D, 3D, and Physicalized Networks}

\author{Michael~J.~McGuffin,
        Ryan~Servera,
        and Marie Forest%
\IEEEcompsocitemizethanks{\IEEEcompsocthanksitem M. McGuffin is with \'{E}cole de technologie sup\'{e}rieure,
Montreal,
Canada.
\IEEEcompsocthanksitem R. Servera was with \'{E}cole
de technologie sup\'{e}rieure,
Montreal, Canada.
\IEEEcompsocthanksitem M. Forest is with \'{E}cole
de technologie sup\'{e}rieure,
Montreal,
Canada.}%
\ifthenelse{\boolean{VersionForJournal}}{
\thanks{Manuscript received Whenuary nn, 2022; revised Whenember nn, 20nn.}
}{} %
}

\ifthenelse{\boolean{VersionForJournal}}{

\markboth{Journal of Blah,~Vol.~V, No.~N, Whenuary~2022}%
{McGuffin \MakeLowercase{\textit{et al.}}: Path Tracing in 2D, 3D, and Physicalized Networks}

}{} %

\IEEEtitleabstractindextext{%
\begin{abstract}
It is common to advise against using 3D
to visualize abstract data such as networks,
however Ware and Mitchell's 2008 study
showed that path tracing in a network is less
error prone in 3D than in 2D.
It is unclear, however, if 3D retains its advantage
when the 2D
presentation of a network is improved using edge-routing,
and when simple interaction techniques for exploring the network
are available.
We address this with two studies of path tracing under new conditions.
The first study was preregistered, involved 34 users,
and compared 2D
and 3D layouts that the user could rotate and move in virtual reality
with a handheld controller.
Error rates were lower in 3D than in 2D,
despite the use of edge-routing
in 2D and the use of mouse-driven interactive highlighting of edges.
The second study involved 12 users and investigated data physicalization,
comparing 3D layouts in virtual reality
versus physical 3D printouts of networks
augmented with a Microsoft HoloLens headset.
No difference was found in error rate,
but users performed a variety of actions with their
fingers
in the physical condition which can inform new interaction
techniques.
\end{abstract}

}

\maketitle

\IEEEdisplaynontitleabstractindextext
\IEEEpeerreviewmaketitle

\IEEEraisesectionheading{\section{Introduction}\label{sec:introduction}}
\IEEEPARstart{A}{dvances} in
virtual reality (VR) and augmented reality (AR) headsets
have fueled interest in 3D graphics for information visualization
and `immersive analytics'
\cite{marriott2018,fonnet2019,ens2021,kraus2022}. %
For datasets with a natural 3D embedding,
such as 3D medical images or 3D models of buildings,
there is clear value in 3D visualization.
On the other hand, for abstract data such as
networks \cite{vonlandesberger2010} %
or multidimensional multivariate data \cite{wong1997},
the use of 3D is often advised
against \cite{munzner2014commentsOn3D}
due to previous studies
that have found 2D to be better (e.g., \cite{sedlmair2013scatterplots,jansen2013evaluating}).
One counter-example is the task
of path tracing in networks,
which was shown in a carefully designed experiment
\cite{ware2008} to be less error-prone when using a 3D layout
with stereo and motion parallax depth cues.
Practical implications remain unclear:
should networks be embedded in 3D?
The lack of clear implications is
partly because
the previous study did not allow the user
to control their view in 3D,
nor leverage interaction with an input device,
nor benefit from modern edge-routing \cite{dwyer2009fast}
in 2D (edges in \cite{ware2008} were simply drawn
as straight line segments, resulting in more occlusion).
We extend this previous work by experimentally comparing path tracing under new conditions that are more relevant to modern VR/AR headsets,
and find that 3D remains advantageous over 2D in terms of error rate. %

\begin{figure*}[!t]
 \centering
 \includegraphics[width=1.9\columnwidth]{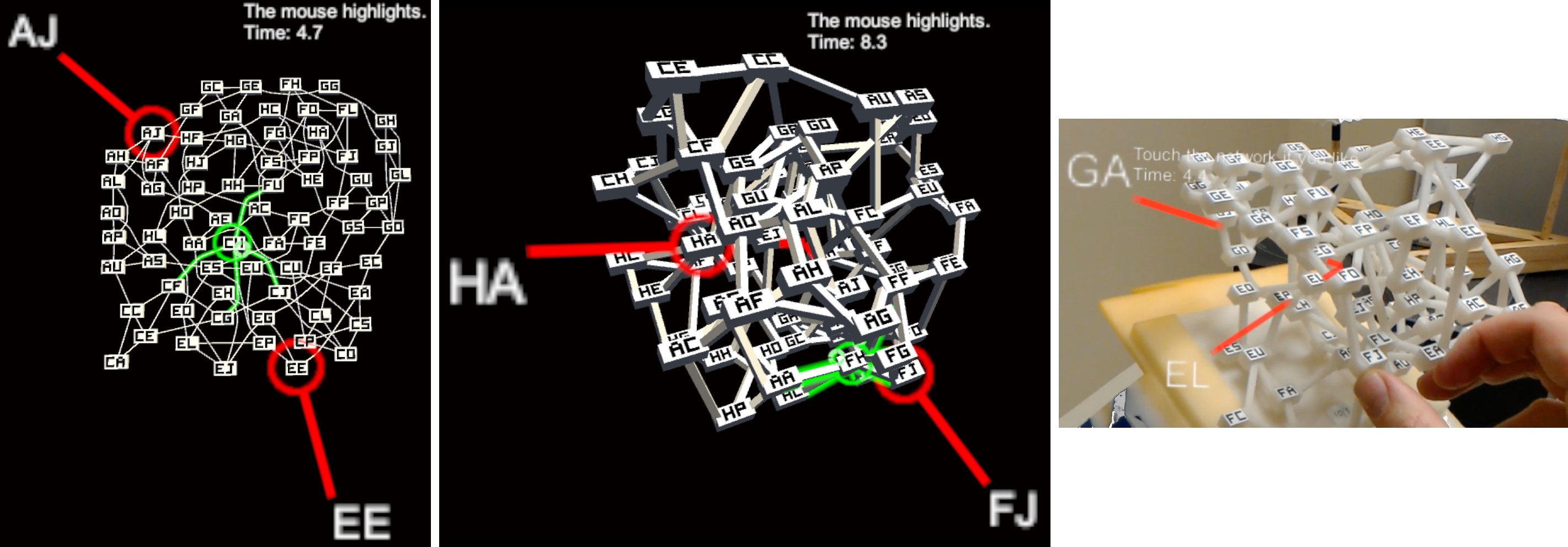}
 \caption{
     In our two studies, users had to find the distance (in edges)
     between the two nodes indicated in red.
     Left and Middle: the \VRTwoHilite\ and \VRThreeHilite\ conditions,
     where a mouse moved a green cursor, highlighting edges incident
     on the node under the cursor.
     Right: the \ARTouch\ condition,
     where the user could touch a 3D printout,
     and a Microsoft HoloLens augmented reality (AR) headset indicated the nodes.
  }
 \label{fig:teaser}
\end{figure*}

Our work is also informed by the recent trend
of physicalization of data \cite{jansen2015phys,dataphyswiki},
e.g., via 3D printing. The ability to touch a
tangible rendering of data
can yield advantages
over an equivalent virtual 3D visualization
\cite{jansen2013evaluating}.
This is likely in part because the user's fingers
can mark elements in a physicalization,
to ``remember'' a location %
and facilitate comparison with other elements. %
To date, however, there have been no empirical evaluations
of physicalizations of networks with 3D layouts.
Physicalizations also open the intriguing possibility of being augmented
with virtual information displayed using an AR headset.
Our 2nd study is the first to experimentally evaluate
a physical network with a 3D layout,
and also the first to use AR to augment physicalized networks, an example of what we call
{\bf augmented physicalization}.

\section{Background}

The choice between visualizing data in 2D and 3D
is most controversial when the data is abstract,
having no intrinsic embedding.
Previous works offer much advice
\cite[Section 3]{munzner2008pitfalls},
\cite{munzner2014commentsOn3D,
brath2014dealwithit,
ware2000attitude,lee2022,
shneiderman2003better}.
Problems with 3D \cite{brath2014dealwithit,sedlmair2013scatterplots}
include occlusion hiding information,
ambiguous depth,
distortion due to perspective,
complex navigation,
and difficulty reading text.
Advantages of 3D include having an additional visual channel for encoding a variable,
ability to have multiple views in 3D space \cite{collins2007vislink,marriott2018},
and depth cues sometimes making it easier to find information \cite{zou2022stereo}.

VR and AR \cite{kim2018} headsets
provide immersion, 3D input,
and enhanced depth cues, in particular head-coupled motion
and stereo disparity \cite{mcintire2014goodbadugly,mcintire2014stereoreview,zou2022stereo}.
Recent uses of these
platforms for information visualization
include
ImAxes \cite{cordeil2017imaxes} and DataHop \cite{hayatpur2020datahop}.
Input to such systems is often via handheld
controllers or whole hands.
Some systems use tangible input devices \cite{cordeil2017tangibleimmersive}
designed to better match
visualization tasks.
Recent examples include
a tangible cutting plane \cite{bach2018hologram}, %
tangible axes \cite{cordeil2020embodied, smiley2021madeAxis},
and a globe of the earth \cite{satriadi2022},
which is both a tangible input device
and a physicalization
\cite{jansen2015phys}
of geographic data.
Other systems
present virtual information on top of
a tangible physicalization {\em without}
using a headset \cite{gillet2005,taher2015emerge,taher2017emerge}.

The next two sections focus on previous
empirical evaluations of
2D vs 3D embeddings of networks,
and visualization vs physicalization of data.

\subsection{Comparing Networks in 2D and 3D}

Network visualization constitutes a large literature
\cite{vonlandesberger2010,dibattista1999,kaufmann2001,yoghourdjian2018,burch2020}. %
Some previous works have evaluated networks
embedded in 3D
\cite{kwon2016, cordeil2017cavehmd, drogemuller2020}
but without focusing on the question of comparing
a flat 2D layout (on a plane) vs fully 3D layout
(with nodes distributed throughout a volume).
In Kwon et al.\ \cite{kwon2016},
the nodes of the network were laid out on a curved surface,
whereas the other works \cite{cordeil2017cavehmd, drogemuller2020}
did not employ a flat 2D layout.
Irani and Ware \cite{irani2003geon} compared
network-like structures rendered with 2D and 3D depth cues,
but always with a flat layout of the nodes.

Other works have compared flat 2D and fully 3D layouts
for tasks related to highlighted subsets of nodes \cite{alper2011stereohiliting}
and counting clusters of nodes \cite{greffard2011community, greffard2014beyond}.

Our work extends previous studies of {\em path tracing}, also called
path finding or path following, where users identify a sequence of nodes.
This is a standard task \cite{lee2006task} with networks,
used in multiple previous studies
\cite{kwon2016,cordeil2017cavehmd,drogemuller2020,james2020,drogemuller2021}
that were not focused on comparing flat 2D vs fully 3D layouts.
Path tracing has also been used to compare monoscopic and stereoscopic
viewing of structures resembling angiograms
\cite{sollenberger1993,vanbeurden2010}.
Studies comparing 2D and 3D layouts for path tracing within networks
are reported in \cite{ware1996evaluating,belcher2003,ware2008}.
Two of these \cite{ware1996evaluating,belcher2003} used networks with random layouts, making them less relevant to real visualizations.
The most recent \cite{ware2008},
summarized below,
is also the most carefully designed.

Unlike our current work, none of the previous studies
involving path tracing
employed edge-routing in their 2D network visualizations.

\subsubsection{Ware and Mitchell (2008)}

Ware and Mitchell \cite{ware2008} report two studies,
and we focus on the first of these, which we abbreviate as W+M.
For each trial, two nodes were highlighted.
Users had to indicate if the shortest path between the two
nodes was 2 or 3 edges, a forced choice response.
There were
5 viewing conditions:
2D layout,
or a 3D layout with \{monoscopic, stereoscopic\} projection
$\times$ \{no motion, motion in the form of automatic rotation at 10$^\circ$ per second\}.
Users could not actively change their view,
either by moving their head
nor through any input device.
Viewing time was limited to 5 seconds per trial
(i.e., a rotation of 50$^\circ$ in the conditions with `motion').
Results showed that the highest error rate occurred in the
2D and 3D monoscopic conditions;
and the lowest error rate was with 3D stereoscopic + motion,
demonstrating an advantage of the fully 3D condition over 2D.

W+M focused on
``visual searches that could be conducted rapidly'' \cite{ware2008}.
Our studies are designed to be more realistic
and relevant to VR/AR headsets.
Our participants
can freely change their view of the network
by moving their head and hand.
In W+M,
the user's field-of-view (FOV) was $\approx$
26$\times$16$^\circ$ per eye,
much smaller than the FOV of the VR headset used
in our Study 1,
and slightly smaller than the AR headset in our Study 2.

In addition, our 2D conditions use 
a state-of-the-art routing
algorithm \cite{dwyer2009fast} (Figures~\ref{fig:teaser}(Left) and Figure~\ref{fig:study1-2D}),
to make better use of space
and reduce ambiguity.
Our experimental task involves paths
that are longer.
Our Study 1 also involves
conditions with interactive highlighting,
double the number of participants of W+M,
and was preregistered (Section~\ref{sec:study1predictions}).

\subsection{Evaluating Data Physicalizations}

Jansen et al.\ \cite{jansen2013evaluating}
evaluated physical barcharts.
Their first study
compared
4 conditions:
2D virtual barcharts,
3D virtual barcharts displayed monoscopically
and stereoscopically (rotation performed with a mouse in both 3D virtual conditions),
and 3D physical barcharts that users could touch.
In terms of time, 2D was the best,
but more interestingly,
3D physical was the 2nd best.
Their second study investigated
why 3D physical might be better than 3D virtual,
comparing 4 conditions:
(1) virtual 3D monoscopic with mouse for rotation,
(2) virtual 3D monoscopic with a prop for more direct rotation,
(3) physical 3D without being allowed to touch,
and (4) physical 3D with touch allowed.
In terms of time,
the 4th condition was best,
and the 3rd condition was 2nd best.

Drogemuller et al. \cite{drogemuller2021} evaluated
networks with a flat, 2D layout,
ranging from 16 to 24 nodes in size,
with %
3 tasks,
comparing 4 conditions:
virtual on-screen (``graphical-only''),
and physical printouts
that could be seen (``visual-only'')
or touched (``haptic-only'') or both (``visual-haptic'').
Users preferred the physical printouts that could be seen
and touched, but within the path tracing task,
no differences are reported in error rates between
graphical-only, visual-only, or visual-haptic.

Our work is the first to empirically
evaluate physicalized networks with 3D layouts.
Also, unlike previous work,
our physicalized networks were augmented with an
AR headset to indicate end-nodes.

\section{Overview of Both Studies}

The following questions motivate our work:
is path tracing easier in networks presented in 3D than in 2D
when edge-routing is used in 2D,
and when the user can interact with the network using a pointing device?
Also, is path tracing easier with a physical 3D representation?

In both our studies, the task was to find the length (between 1 and 5 edges) of a shortest path between two end-nodes indicated by the system.
The user's non-dominant hand (NDH) held and repositioned the network,
because this matches the use of the NDH in the kinematic chain model \cite{guiard1987},
and because there is some evidence that rotation
via a handheld prop
is superior to using a mouse for the same purpose \cite{jansen2013evaluating},
and because it provides an easy-to-understand way to
simultaneously
pan and zoom within a 2D layout,
by simply translating the layout sideways or holding it closer or father away.
In addition, in some conditions, the user's dominant hand (DH) could
move a mouse cursor over nodes (causing incident edges to highlight)
or touch a physical 3D printout of the network.

In all conditions of both studies, the NDH
activated a trigger button to open a radial button
to provide the user's answer from 1 to 5.
The use of the NDH in this way allowed the user to complete each trial without
the ``homing time'' of moving a hand back and forth between two places.
(Had we instead used the DH to open the radial menu, then the \ARTouch\ condition in Study 2
would have required having the user move their DH between the physical network and a button to open
the menu.)
A radial menu was used so that every answer would take the same amount of time to select.

A single set of networks was used for both studies,
from which
networks were randomly chosen
for each condition and each user.

\subsection{Network Size, Topology and Layout}\label{sec:networks}

We generated 10 networks.
For each network, we computed its layout in 3D,
and projected the 3D node positions down to a plane to obtain a layout in 2D.
Each network can be displayed in virtual 2D or 3D,
and was also
3D printed using stereolithography (SLA) with a white plastic.

Each network has 70 nodes
and 140 edges (hence an average degree of 4),
and was generated with a Watts-Strogatz \cite{watts1998} small-world synthesis algorithm.
The algorithm begins by constructing a regular ring lattice of 70 nodes
each with degree 4.
Each edge is then randomly rewired with 20\% probability.
The average degree distribution that resulted over the 10 networks
was: 1, 15.7, 38.2, 12.9, 1.8, and 0.4 nodes
of degree 2 through 7, respectively.

Each of the 70 nodes was assigned a unique 2-character string label such as ``AA'', ``FE'', or ``HL''.

Layout of nodes was performed in two passes.
The first pass uses stress majorization
(equation 12 in \cite{gansner2004}) to position the nodes in 3D.
Projecting node positions down to a 2D plane results in overlap between labels,
hence a second pass applies repulsive forces between nodes whose labels overlap in the 2D plane, pushing nodes away from each other in the horizontal plane.
The new positions are saved in both 2D and 3D.
Thus, the 2D layout uses the projected coordinates of the 3D layout.

Next we compute the layout of edges.
In the 3D case, each node is modeled as an
elongated box (7$\times$7$\times$15 mm)
with a text label on one side,
and each edge is modeled as a single segment (3~mm thick),
with each edge's endpoint connected to the center or extremity
of the node's box in such a way as to avoid
extending through the labeled face of the box.
In 2D, nodes are 7$\times$11~mm rectangles,
and each edge is a multi-segment polygonal line (0.5 mm thick),
whose layout is computed using
the MSAGL (Microsoft Automatic Graph Layout)
library \cite{msagl}
based on \cite{dwyer2009fast}.

The networks displayed in 3D virtual form have the same geometry
as the physically printed networks: the same node dimensions, same edge thickness, same color (white),
and same font used for labels.
The 2D virtual form also uses the same color and font.

The number of nodes for the networks
was chosen based on physical limits of common 3D printers.
First, we wanted networks that would fit within a
6$\times$6$\times$6 inch volume,
to accommodate lower-end printers.
In addition, a 6-inch width
fits within the HoloLen's FOV
at a distance of 30~cm, well within arm's reach.
Second, we wanted edges to be 3~mm thick to avoid fragility.
Third, we wanted nodes big
enough to accommodate embossed text labels
that would be clear even on lower-end
FDM
(Fused Deposition Modeling) printers.
We implemented a custom font (each character a 5$\times$4 bitmap)
allowing us to print text labels on nodes 7~mm high
with 1~mm stroke thickness.
We found that 70 nodes resulted in a network
of reasonable complexity that fit the size constraints.
Testing revealed that embossed text on mono-color 3D printouts
is both difficult to read and to paint,
and because multi-color 3D printing is
much more expensive, we finally used 2D printed stickers
with the same custom font to label each node.

\section{Study 1: 2D and 3D Virtual}

Study 1 was preregistered \cite{mcguffin2022prereg},
done in VR,
and crossed the dimensionality \{2D, 3D\} of the network layout
with the use of a mouse to highlight edges.

\subsection{Main Conditions} %

\begin{figure}[tb]
 \centering
 \includegraphics[width=\columnwidth]{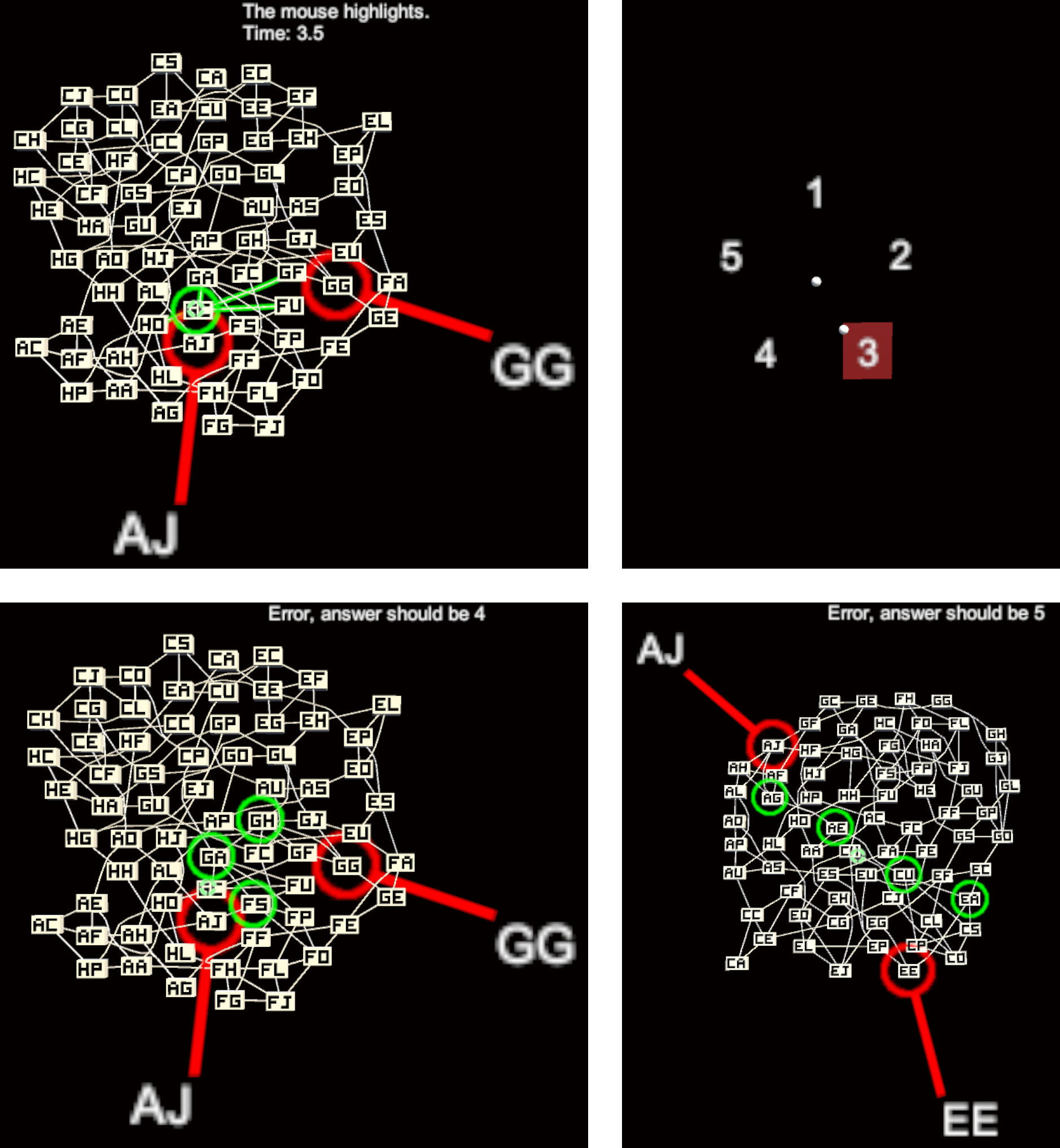}
 \caption{
    In study 1, in the \VRTwoHilite\ condition,
    the user can move the mouse cursor over a node, causing
    incident edges to highlight in green (Top Left).
    Once the user has determined the distance in edges between the
    two nodes indicated in red (AJ and GG), they hold down the trigger
    button with their non-dominant hand (NDH), causing a radial menu to appear
    (Top Right). By tilting their held down and right, they select
    ``3'' in the menu, and then release the trigger button
    to complete their answer. In this example, their answer is wrong,
    so the system displays error feedback:
    a correct shortest path of length 4 is highlighted
    (Bottom Left). Another example of error feedback
    (Bottom Right) is for the network
    shown in Figure~\ref{fig:teaser}(Left), where the shortest path has length 5.
  }
 \label{fig:study1-2D}
\end{figure}

\begin{figure}[tb]
 \centering
 \includegraphics[width=\columnwidth]{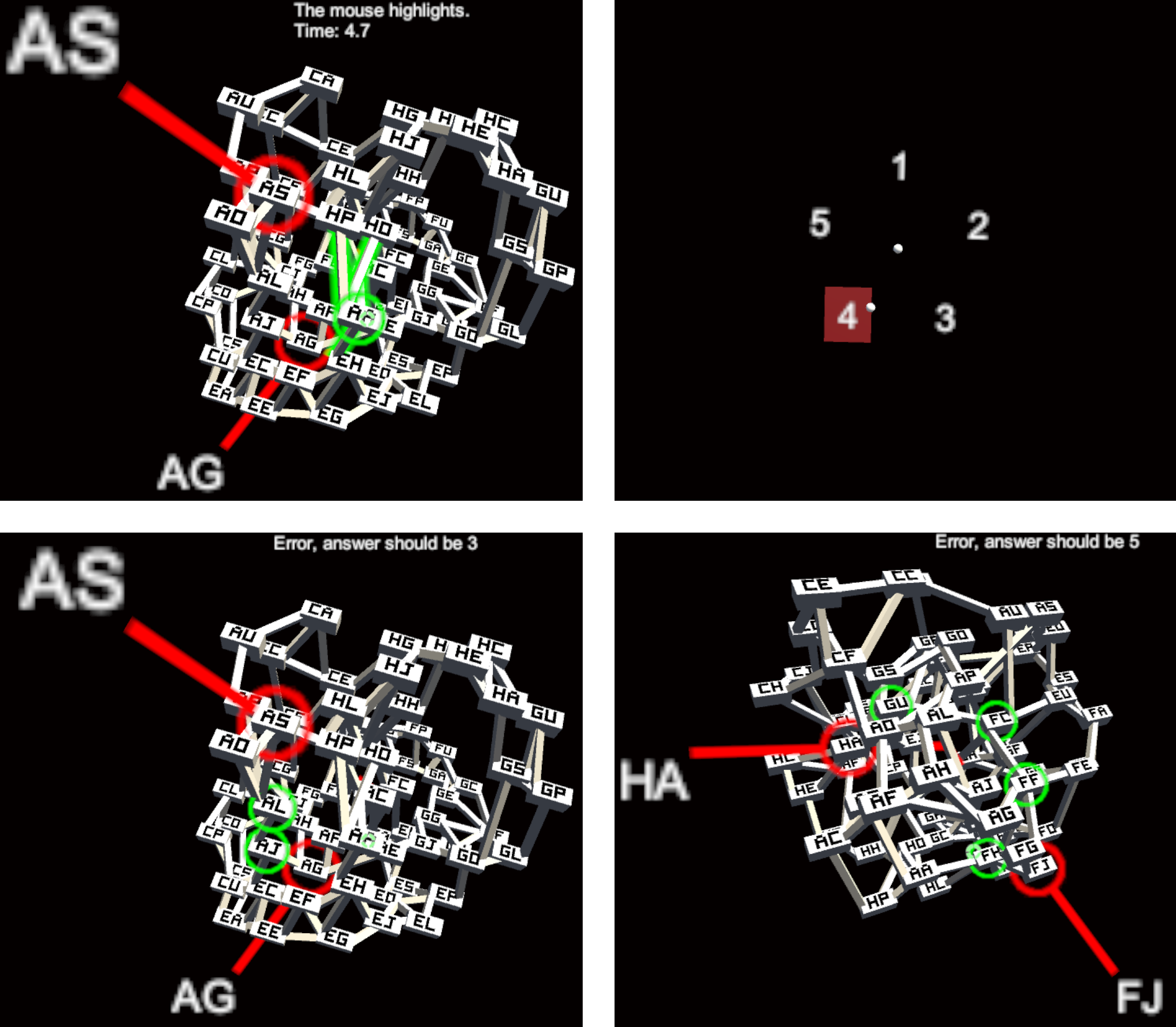}
 \caption{
    Study 1, the \VRThreeHilite\ condition.
    The first three images show a trial ending with error feedback
    displaying a correct shortest path of length 3 (Bottom Left).
    The last image (Bottom Right) shows a correct shortest path of length 5 for the network shown in Figure~\ref{fig:teaser}(Middle).
  }
 \label{fig:study1-3D}
\end{figure}

The previous W+M study \cite{ware2008}
found that 3D outperformed 2D,
but it is plausible that this could change
if the 2D condition is improved with edge-routing,
or if the user can use a pointing device for simple interaction with
the network.
The simplest interaction we could think of
that might help with path tracing is for the edges incident on a node
to highlight when the user hovers over that node with a
pointing device.
This led to our choice of conditions for Study 1.
The independent variable \MainCondition\ has
four possible values:
\begin{itemize}
    \item \VRTwo: Virtual network displayed in VR with 2D layout. The non-dominant hand (NDH) holds a controller to position the network with 6 degrees of freedom (DoF).
    \item \VRTwoHilite: Same as previous, but
              with a mouse in the dominant hand (DH) used to highlight edges.
    \item \VRThree: Virtual network displayed in VR with 3D layout.
        The NDH holds a controller to position the network with 6 DoF.
    \item \VRThreeHilite: Same as previous, but
              with a mouse in the DH used to highlight edges.
\end{itemize}

We decided to not compare with 2D conditions on a desktop screen
(without VR headset) because this would have introduced confounds
in having different display hardware,
and also potentially different input devices.
The VR handheld controller in the user's NDH provides
an easy-to-understand and quick way to
reposition a 3D network layout, and also
simultaneously pan and zoom
when examining a 2D network layout,
and there is no equivalent NDH input on standard
desktop PCs.

\subsection{Task}  %

The experiment consisted of a sequence of trials
where the task
required the user to find the length, in edges,
of a shortest path between two nodes (the path's {\em end-nodes}) in the network.
The independent variable \PathLength\ ranged from 1 to 5,
where 1 means the end-nodes are neighbors.
The shortest path was not necessarily unique.

At the start of each trial, the two end-nodes were indicated
with red callout line segments,
as well as with red circular rings.
The user then examined the network by repositioning and rotating it with their
NDH.
Bringing the network closer to their eyes allowed them
to effectively ``zoom in'' to see more detail.
In some conditions (\VRTwoHilite\ and \VRThreeHilite), the user could also move a mouse with their DH, causing a cursor to hover over different nodes.
Whichever node was under the mouse cursor was highlighted
in green,
as were all the edges incident on that node. %
Users were told that using the mouse in these conditions was not mandatory,
but that it might allow them to more easily find the shortest
path and confirm its length.
The mouse buttons served no purpose.

Once the user thought they knew the answer, they pressed a trigger button using
the index finger of their NDH to open up a radial menu containing the answers
1 through 5 (Figures~\ref{fig:study1-2D}(Top Right) and \ref{fig:study1-3D}(Top Right)).
To select within this menu, the user tilted their head slightly
in the direction of the desired answer and released the trigger button.
Releasing the trigger button without tilting their head dismissed the radial menu
and allowed the trial to continue.
The trial ended only when the user made a selection within the menu.
A text message appeared immediately after to inform the user if their answer
was correct or not, and in the latter case,
the text also indicated the correct answer,
and the system highlighted a shortest path
(Figures~\ref{fig:study1-2D}(Bottom) and \ref{fig:study1-3D}(Bottom)).
The system then moved on to the next trial.

Users were aware that they could take their time during the warmup trials
at the start of each condition,
but after these warmup trials,
they were instructed to complete trials as quickly as possible with no errors.
The following was also explained to each user.
The correct path length is always at most 5,
and, sometimes, the shortest path is quite difficult to find.
After 20 seconds into a trial, the text instructions displayed by the headset turn red. Once this happens, the user is free to continue searching for a shortest path if they so desire, but a reasonable strategy after 20 seconds would also be to simply estimate an answer such as 5, or perhaps 4, even if the user has not found a path of that length.
(This was explained to avoid having users spend too much time searching for difficult paths.)
On the other hand, if a user answers quickly and incorrectly,
the system imposes a ``punishment'' of a delay of up to 15 seconds
before proceeding to the next trial.
(If $t$ was the time in seconds taken by the user to give an incorrect answer,
the precise delay imposed after the trial
was $\max(\min(20-t,15),5)$, i.e., a decreasing ramp function
clamped between 15 and 5 seconds.)
This disincentivizes a user from answering sloppily to complete the experiment faster.

The software asked users to take breaks between conditions.
Users could indicate that they were ready to proceed by pressing a key
on a keypad with their DH.
A Lego brick attached to the key made it easier to feel when the user
was wearing the VR headset.

\subsection{Pilot and Predictions}\label{sec:study1predictions} %

A pilot was performed with 6 users,
after which minor tuning to the protocol was made,
and a preregistration\cite{mcguffin2022prereg}
was archived to declare the number of users to recruit,
criteria for including participants,
predictions to test,
and the R script for plotting data and testing predictions.

Two predictions were preregistered:
first, that the error rate (averaged over trials of \PathLength\ 2, 3 and 4, and averaged over conditions with and without the mouse)
would be smaller in 3D than in 2D;
and second, that the error rate (averaged over trials of \PathLength\ 2, 3 and 4, and averaged over 2D and 3D conditions)
would be smaller with the mouse than without the mouse.
The R script tests each of these predictions by computing a single error rate for each
user and each subset of conditions, not including warmup trials,
and then performing a paired sample $t$-test. %

The W+M study \cite{ware2008} found that 3D yielded
a smaller error rate, possibly because
stereo and motion disambiguate edges.
A second mechanism that could play a role is
that shortest paths in 3D tend to follow a more straight line
(hence, are easier to perceive)
than in 2D.
This 2nd mechanism may not have been at play in W+M
because they
``selected paths in such a way that the mean Euclidean distance between start nodes and end nodes was the same'' \cite{ware2008},
regardless of whether the path was 2 or 3 edges long.
However, both mechanisms could benefit 3D in our study,
since we do not hold this Euclidean distance constant.
The edge-routing in our 2D layouts can also make 2D layouts
easier to read,
but the paths are still ``less straight'' than in 3D.
The reason our predictions about error rates exclude
\PathLength\ 1 and 5 is that those cases tend to be easier for users: in the case of 1, the nodes are often clearly adjacent,
and in the case of 5, the user knows that the length cannot be greater than
5 and can therefore guess a length of 5 when the path is difficult to find.

\subsection{Mouse Cursor in 2D and 3D}  %

To allow the user to hover over a node for interactive highlighting,
we wanted to use the same pointing device in the 2D and 3D conditions,
for simplicity and consistency across conditions.
Although a 6 DoF handheld controller could have been used for pointing,
the mouse is very often used for raycast pointing in 3D,
whereas controllers are rarely used for pointing in 2D.
Furthermore, a 2D mouse may be easier and less tiring to control
than a handheld controller, because the user's arm can partially rest on the
desk, and the depth dimension is automatically handled by the software.
We therefore implemented a variant of raycast pointing with a 2D mouse.

In our variant of raycast pointing,
we wanted the user
to be able to position the mouse cursor over a node with their DH,
and then move or rotate the network with their NDH while the mouse cursor
remains `stuck' on the same node.
Therefore, the mouse cursor's position is stored in the network's local 3D space,
and the DH applies relative displacements.
Whenever the DH moves the mouse, a proportional translation is applied to the mouse
cursor parallel to the camera plane (i.e., the plane perpendicular to the camera's forward direction).
After applying this translation, a ray is cast from the camera position through the cursor,
and if this ray encounters one or more nodes, our software finds the intersection between the ray
and the node closest to the camera, and moves the cursor to that intersection.
The user therefore sees the cursor automatically jump forward or backward
to stick to the nearest node, but these jumps only happen if the mouse is being moved.
If the user is only using their NDH to reposition or rotate the network,
no automatic jumps happen, and the cursor retains its position in the network's local space.

Like EZCursorVR \cite{ramcharitar2018},
the size of our mouse cursor is scaled to be bigger when the cursor
is farther from the camera, so that the projected size of the cursor on the camera plane
appears constant.

\subsection{Hardware} \label{sec:study1hardware} %

\begin{figure}[tb]
 \centering
 \includegraphics[width=\columnwidth]{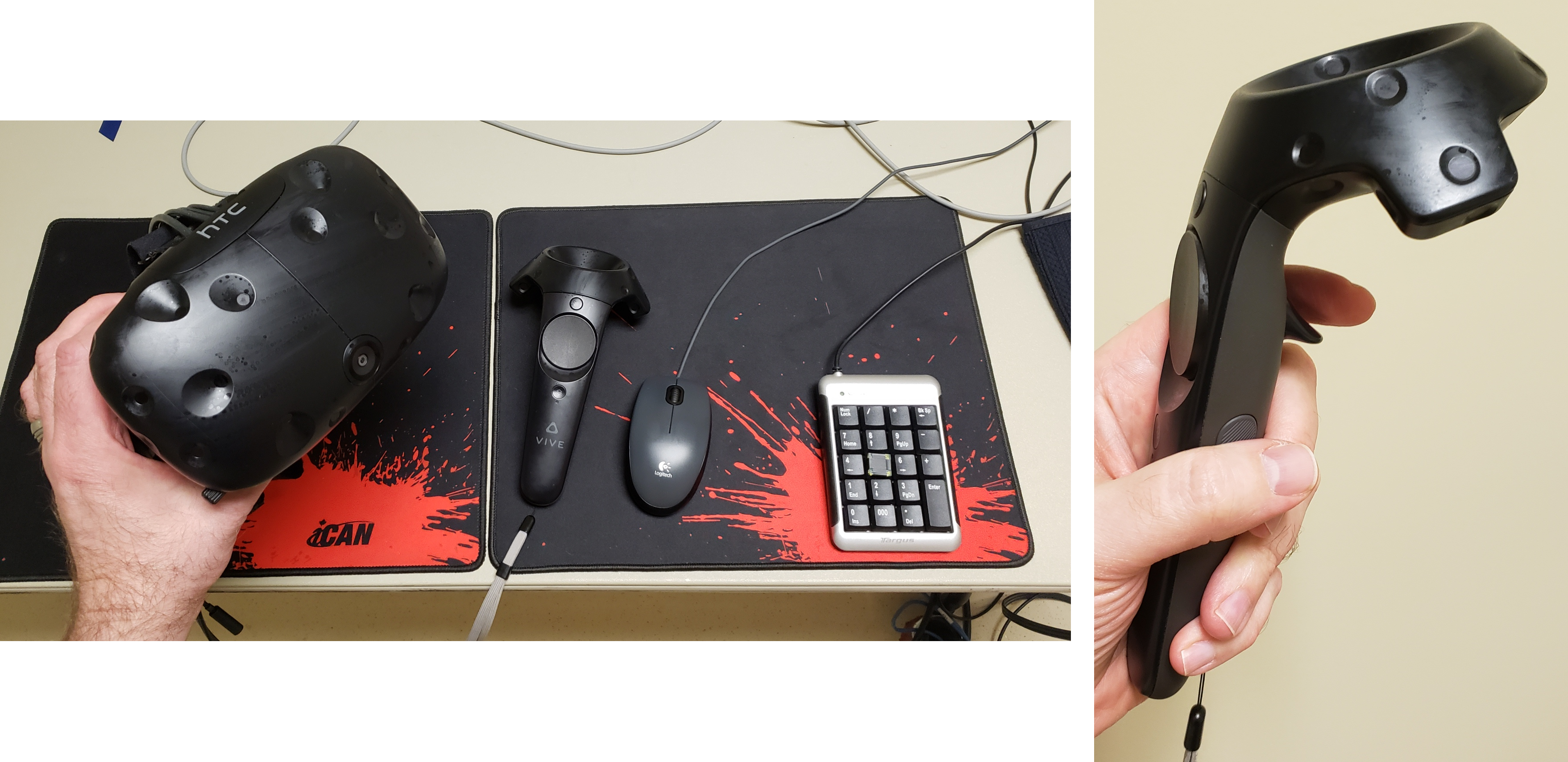}
 \caption{
    The equipment for Study 1:
    HTC Vive headset and controller (held in the NDH),
    mouse (for the DH)
    and keypad (to advance to the next trial after a break).
  }
 \label{fig:study1hardware}
\end{figure}

We used an HTC Vive headset,
which has a 2160$\times$1200 resolution (1080$\times$1200 per eye)
and $\approx$110$^\circ$
FOV. %
We measured a framerate of 90 fps.
The handheld controller was held in the user's NDH.
The headset was connected to a PC with an Intel i7-8700K 6-core CPU at 4.7 GHz,
water cooling,
and Nvidia GeForce GTX 1080 Ti GPU.
A Logitech M100 mouse
(with default acceleration settings in Microsoft Windows 10)
on a gaming mouse pad was held in the user's DH.

\subsection{Measurement of rotation} \label{sec:rotationCalculation}

In the 3D conditions, we %
measured how much the user looked at the network from different points of view. %
Let $H_t$ be the position of the user's head at time $t$,
and let $N_t$ be the pose (position and orientation) of the network at time $t$.
We define a direction vector $d = d(H,N)$ as a function of $H$ and $N$,
where $d$ points from the network to the head, and where the components of $d$
are computed in the local space of the network
(i.e., a change in the position of the head, or in the position of the network,
or in the orientation of the network, will each change the components of $d$).
We compute $d_t = d(H_t,N_t)$ for each frame during a trial,
and at the end of the trial, we compute the mean direction $\bar{d}$,
and compute and record the standard deviation of the angles between all the $d_t$ and $\bar{d}$.
This standard deviation is an overall measure of how much the user looked at the network from
different points of view.

We also computed how much the user rotated the network with their hand,
versus how much they moved their head to look at the network from different points of view,
expressed as two percentages that sum to 100\%.
To compute these percentages,
we compute the directions
$d_{t,H} = d(H_t,N_{t-1})$ and $d_{t,N} = d(H_{t-1},N_t)$
that would have resulted if only the head, or only the network, respectively,
had moved.
We find the angle $\alpha_H$ between $d_{t-1}$ and $d_{t,H}$,
and the angle $\alpha_N$ between $d_{t-1}$ and $d_{t,N}$,
and then define the contribution of the head motion to the rotation as $\alpha_H / ( \alpha_H + \alpha_N )$,
and the contribution of the network's motion as $\alpha_N / ( \alpha_H + \alpha_N )$.
These fractions are computed for each frame of the trial (except the first frame),
averaged over the entire trial,
and recorded as percentages.

\subsection{Protocol} \label{sec:study1protocol} %

Equipment was disinfected prior to each user session.
At the start of each session,
after signing a consent form,
users had their
interpupillary distance (IPD) measured, %
as well as their
stereo acuity,
which was assessed
using the `circle test' of the FLY stereo acuity test
by Vision Assessment Corporation. %
(This resulted in a score on a scale of 10, corresponding to
a disparity of
400, 200, 160, 100, 63, 50, 40, 32, 25, or 20 seconds of arc.)
Users then filled out a pre-questionnaire,
and were also shown several printouts of example trials,
with conditions in random order,
to explain the task and test their understanding of
path length.
The equipment for the experiment was then explained to the user,
the headset was adjusted for comfort, and the IPD of the headset
was set to the value measured earlier.
After the trials were completed, a post-questionnaire was filled out.

\subsection{Users} %

A sample size of 34 users was chosen in the preregistration,
not including the pilot participants. This number was chosen to achieve a power of 0.8 at $\alpha=0.05$ for a medium effect size of 0.5, as calculated using the G*Power software\cite{gpower} and also with an online calculator \cite{dhand2014}.

Of the 34 users,
24 were male, 10 female;
28 right handed, 4 left handed, 2 ambidextrous, but all with a habit of using the mouse with their right hand;
age 19 to 45 years (average 24.7);
IPD 54 to 70mm (average 63.7);
stereoacuity test scores 3/10 to 10/10 (average 8.5).

\subsection{Design}  %

Each user experienced the
4 levels of \MainCondition\ \{\VRTwo, \VRTwoHilite, \VRThree, \VRThreeHilite\}
in random order.  For each \MainCondition,
the user performed 10 warmup trials in random order
(5 levels of \PathLength\ $\times$ 2 repetitions)
with the warmup network,
followed by 20 trials in random order
(5 levels of \PathLength\ $\times$ 4 repetitions)
with another network chosen at random,
followed by another 20 trials with another network.
There were a total of
34 users
$\times$ 4 levels of \MainCondition\ 
$\times$ 2 networks
$\times$ 5 levels of \PathLength\
$\times$ 4 repetitions per trial
= 5440 trials, not counting warmup trials and not counting pilot data.
Each session with a user lasted $\approx$ 1.5 hours.

\subsection{Results} %

For these results,
advice was adapted
from Dragicevic \cite{dragicevic2016fair}.
Although we report some $p$ values,
we do not emphasize null hypothesis significance testing (NHST) %
as we wish to avoid misleading, dichotomous thinking
(``Tip 25'' in \cite{dragicevic2016fair}).
We present effect sizes visually
and with CIs (Tips 15, 16),
where the CIs are computed using one (averaged) value for
each (user, condition) pair (Tip 9);
and we clearly distinguish between pre-experiment predictions and
post-hoc exploratory data analysis,
to avoid
HARKing (Hypothesizing After the Results are Known) and $p$-hacking.

All CIs are 95\%.
The CIs for error rates were calculated using
bootstrapping, to prevent them from falling outside the [0,100\%] range.
For other variables such as time, CIs were calculated using
the $t$-distribution.

\begin{figure}[tb]
 \centering
 \includegraphics[width=\columnwidth]{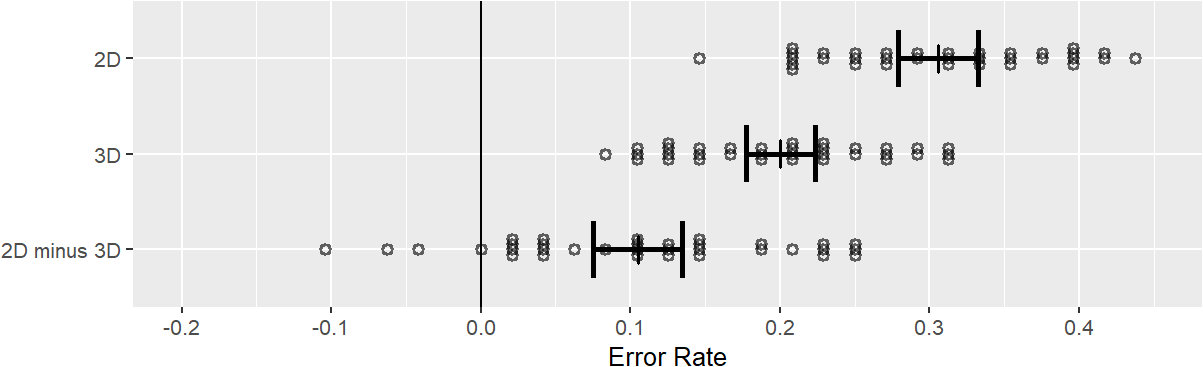}
 \caption{
    Error rates in study 1, %
    excluding \PathLength\ 1 and 5.
    ``2D'' is the union of \VRTwo\ and \VRTwoHilite;
    ``3D'' is the union of \VRThree\ and \VRThreeHilite.
    Each dot is the average for one user,
    and the bars show 95\% confidence intervals (CIs),
    computed from the 34 users.
    A paired $t$-test yields $p<0.0000005$.
    The 3D conditions resulted in a lower error rate than 2D,
    confirming the first of our preregistered predictions.
  }
 \label{fig:results-study1-prediction1}
\end{figure}

\begin{figure}[tb]
 \centering
 \includegraphics[width=\columnwidth]{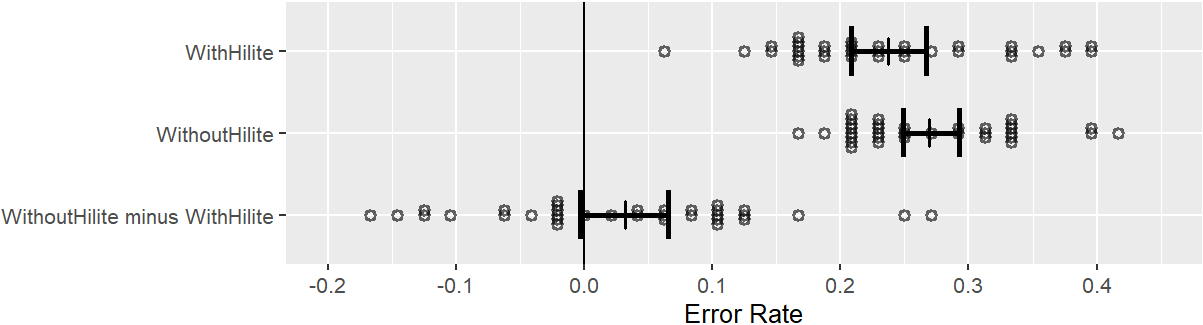}
 \caption{
    Error rates in study 1, %
    excluding \PathLength\ 1 and 5.
    ``WithHilite'' is the union of \VRTwoHilite\ and \VRThreeHilite;
    ``WithoutHilite'' is the union of \VRTwo\ and \VRThree.
    A paired $t$-test yields $p<0.09$,
    providing limited evidence of our second preregistered prediction,
    that the mouse improves error rates.
  }
 \label{fig:results-study1-prediction2}
\end{figure}

Figures~\ref{fig:results-study1-prediction1} and
\ref{fig:results-study1-prediction2}
show the results of testing the preregistered predictions.
The fact that the zero line falls far outside
the CI of the difference in the first figure,
and barely intersects the CI of the difference in the second
figure,
is reflected by the very small $p$ value in the first case
and a $p$ value somewhat larger than 0.05 in the second case.
We thus have strong evidence that 3D results in a lower error rate,
and limited evidence that the error rate is reduced by
interactive highlighting of edges with the mouse.

\subsubsection{Subjective Feedback and Exploratory Data Analysis} %

\begin{figure}[tb]
 \centering
 \includegraphics[width=\columnwidth]{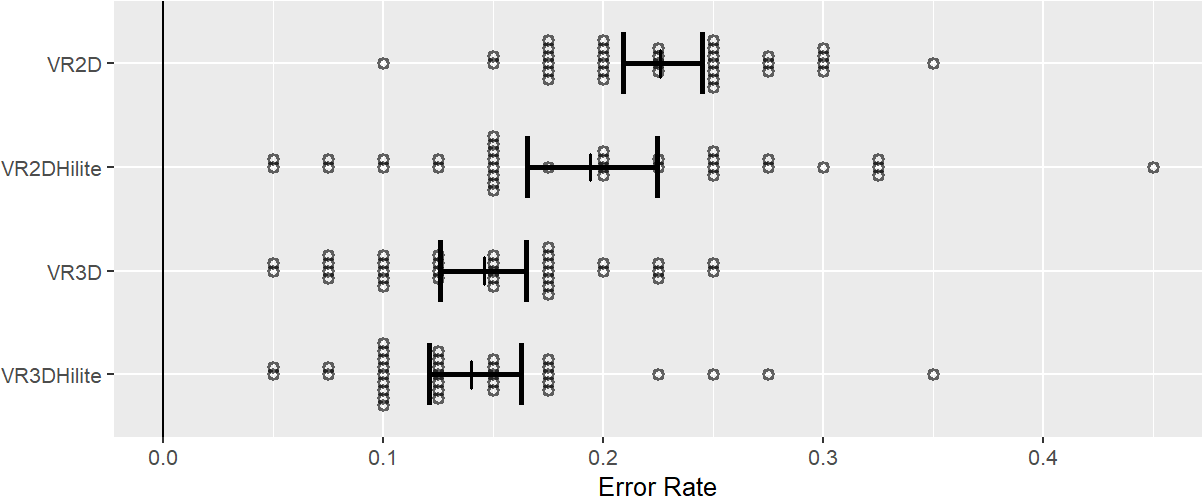}
 \caption{
    Error rates in study 1 by \MainCondition,
    including all \PathLength\ values 1-5.
  }
 \label{fig:results-study1-errorRate2}
\end{figure}

\begin{figure}[tb]
 \centering
 \includegraphics[width=\columnwidth]{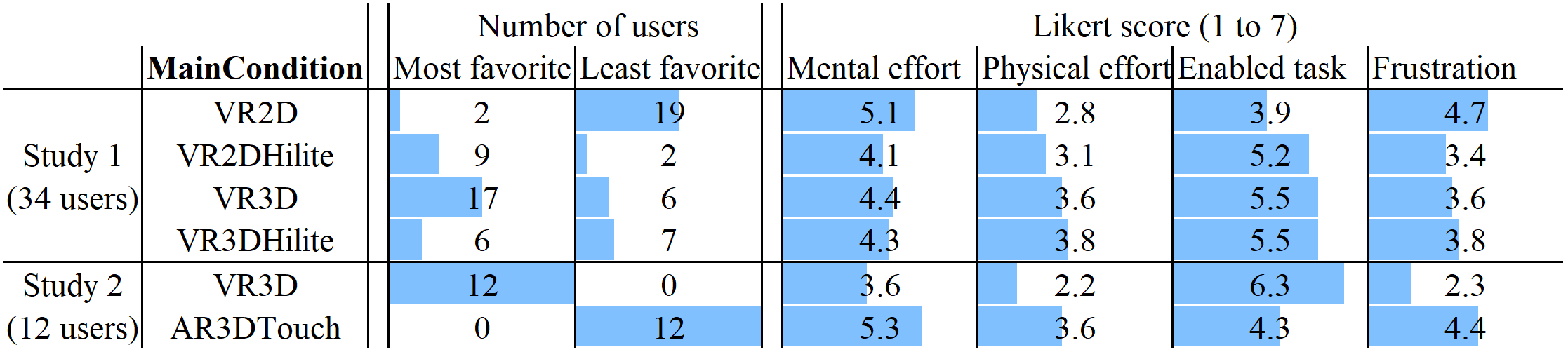}
 \caption{
    Subjective results of each study.
  }
 \label{fig:subjective}
\end{figure}

In subjective comments,
10 out of the 34 users described 3D positively as compared to 2D, saying that 3D makes the task easier, simpler, less confusing, fun, making pathways clearer, only requiring rotation.  One of these users mentioned longer paths as being clearer in 3D than in 2D.

Figure~\ref{fig:results-study1-errorRate2} shows
error rates by \MainCondition, suggesting
that interactive highlighting of edges
with the mouse helped in 2D but not in 3D.
This is further supported by comments made by users
and by the subjective results in
Figure~\ref{fig:subjective}.
(For ease of comparison, the results
of Study 2, discussed later,
are presented alongside several figures.)
There, we see that \VRTwo\ was
the least favorite condition,
and that \VRTwoHilite\ required
less mental effort,
produced less frustration,
and better enabled the task.
However, \VRThree\ was the most favorite
condition, not \VRThreeHilite.
14 out of the 34 users described the highlighting of edges with the mouse in 3D (\VRThreeHilite) in negative terms, such as being not intuitive, requiring extra motion and time, or difficult to position the cursor in the depth dimension.
Note that users were given no explanation of
how the mouse worked in 3D,
and it is possible that several users were confused by
it because they were moving their NDH and DH at the same time.
Despite this, Figure~\ref{fig:subjective} also shows that
6/34 = 18\% of users chose \VRThreeHilite\ as their favorite condition,
and 8 users described it as helpful, useful, faster, requiring less rotation of the network, and making pathways more visible.

\begin{figure}[tb]
 \centering
 \includegraphics[width=\columnwidth]{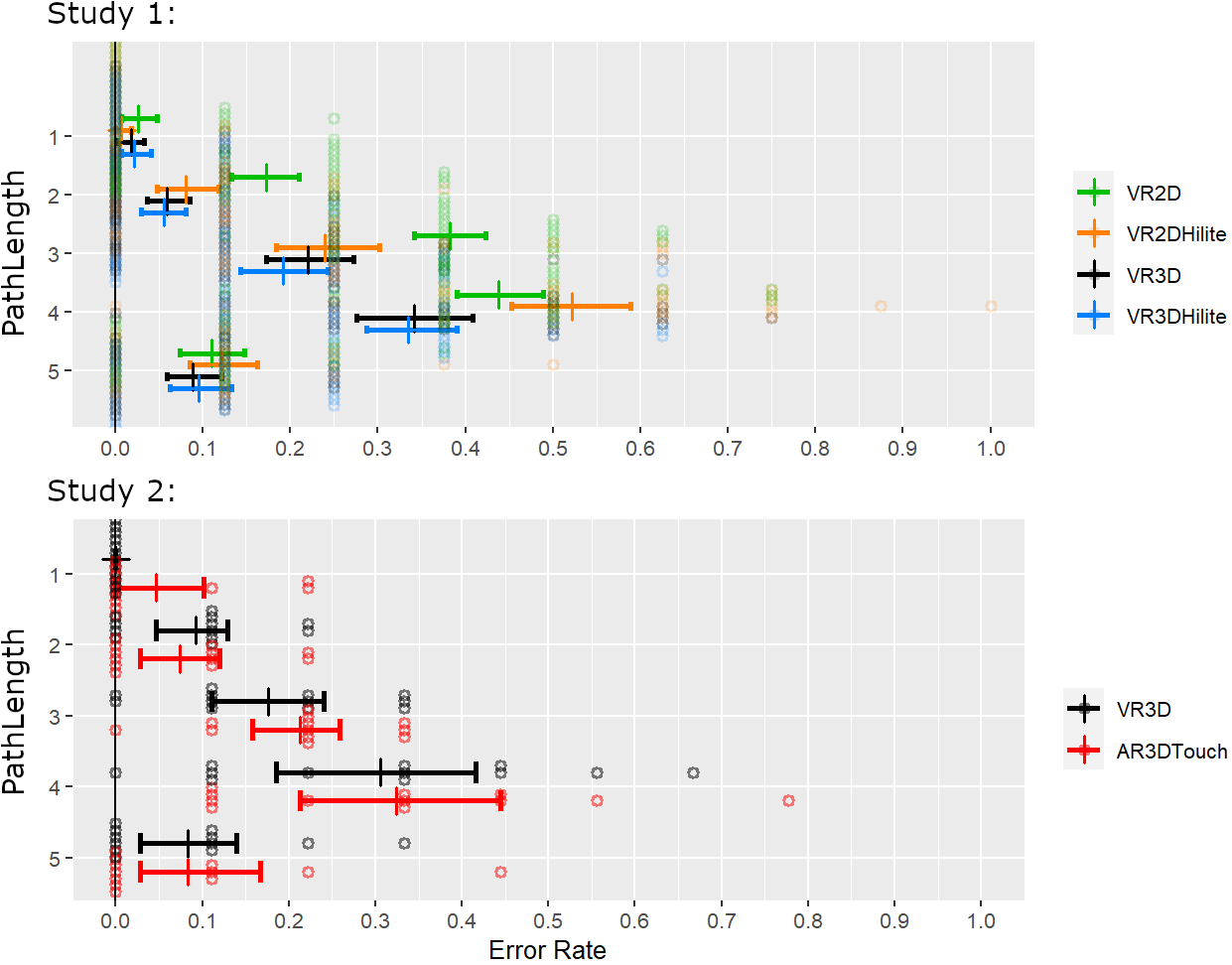}
 \caption{
    Error rates in more detail.
    Each dot is the average for one user,
    and the bars show 95\% CIs,
    computed from the 34 or 12 users, in studies 1 and 2, respectively.
    In the 2D conditions (green and orange) of study 1, notice that edge highlighting with the mouse appears to have helped for \PathLength\ 2 and 3 but not 4.
  }
 \label{fig:results-errorRate}
\end{figure}

\begin{figure}[tb]
 \centering
 \includegraphics[width=\columnwidth]{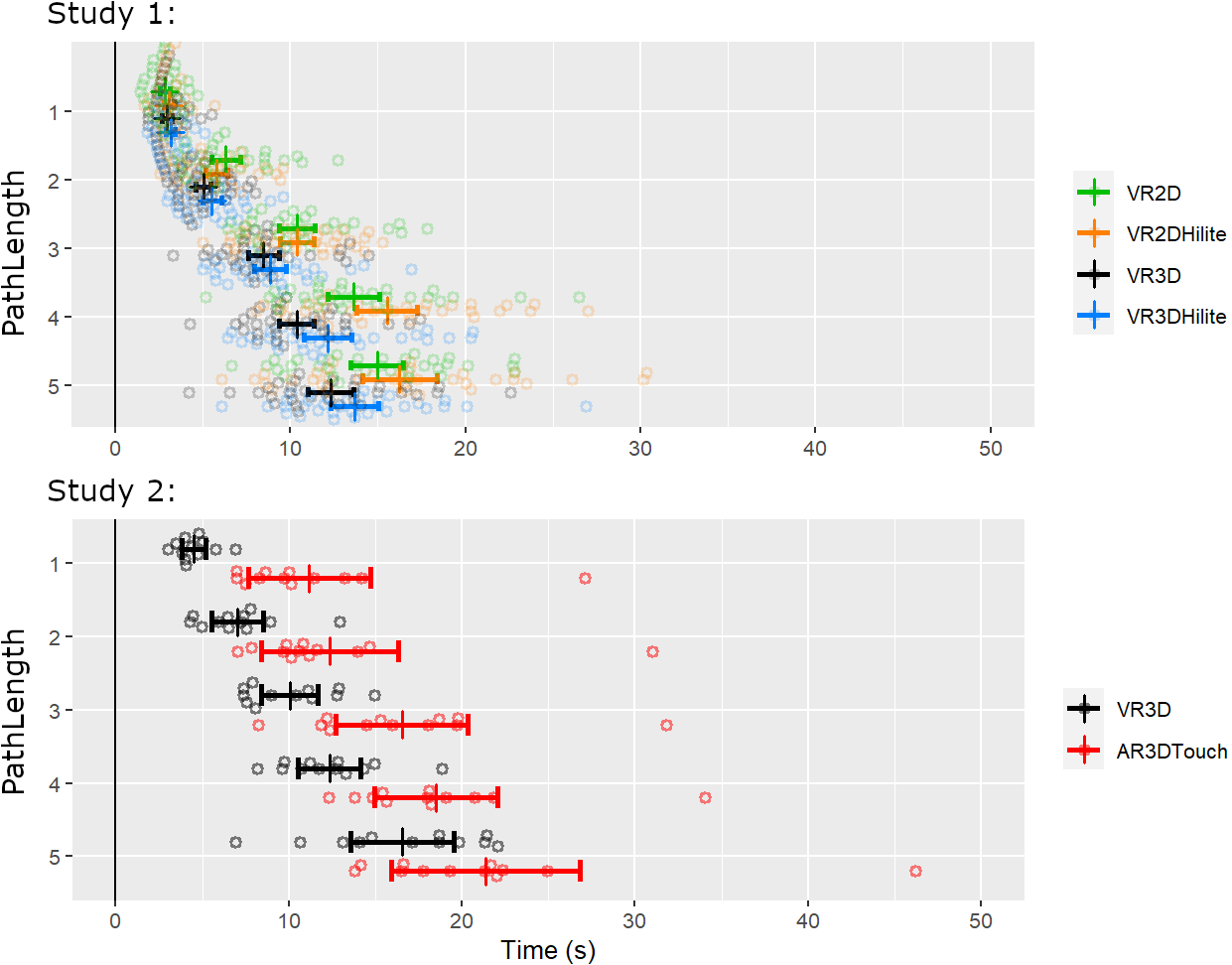}
 \caption{
    Duration of trials.
  }
 \label{fig:results-time}
\end{figure}

Figures~\ref{fig:results-errorRate}
and \ref{fig:results-time}
present the error rates and
times in more detail.
Notice that both the times and error rates appear
smaller in 3D than in 2D.
In Figure~\ref{fig:results-errorRate},
if we examine the 2D conditions in Study 1,
it seems that the mouse helped with \PathLength\ 2 and 3
but hindered performance with \PathLength\ 4.
This may be because the interactive highlighting sometimes
misled %
users into following suboptimal paths.
This is partially supported by
comments made by
3 users, who talked about the edge highlighting causing them to focus more on those edges, inducing a different ``mental exercise'' than without edge highlighting, and encouraging the user to explore the network by ``testing'' different edges.

\begin{figure}[tb]
 \centering
 \includegraphics[width=\columnwidth]{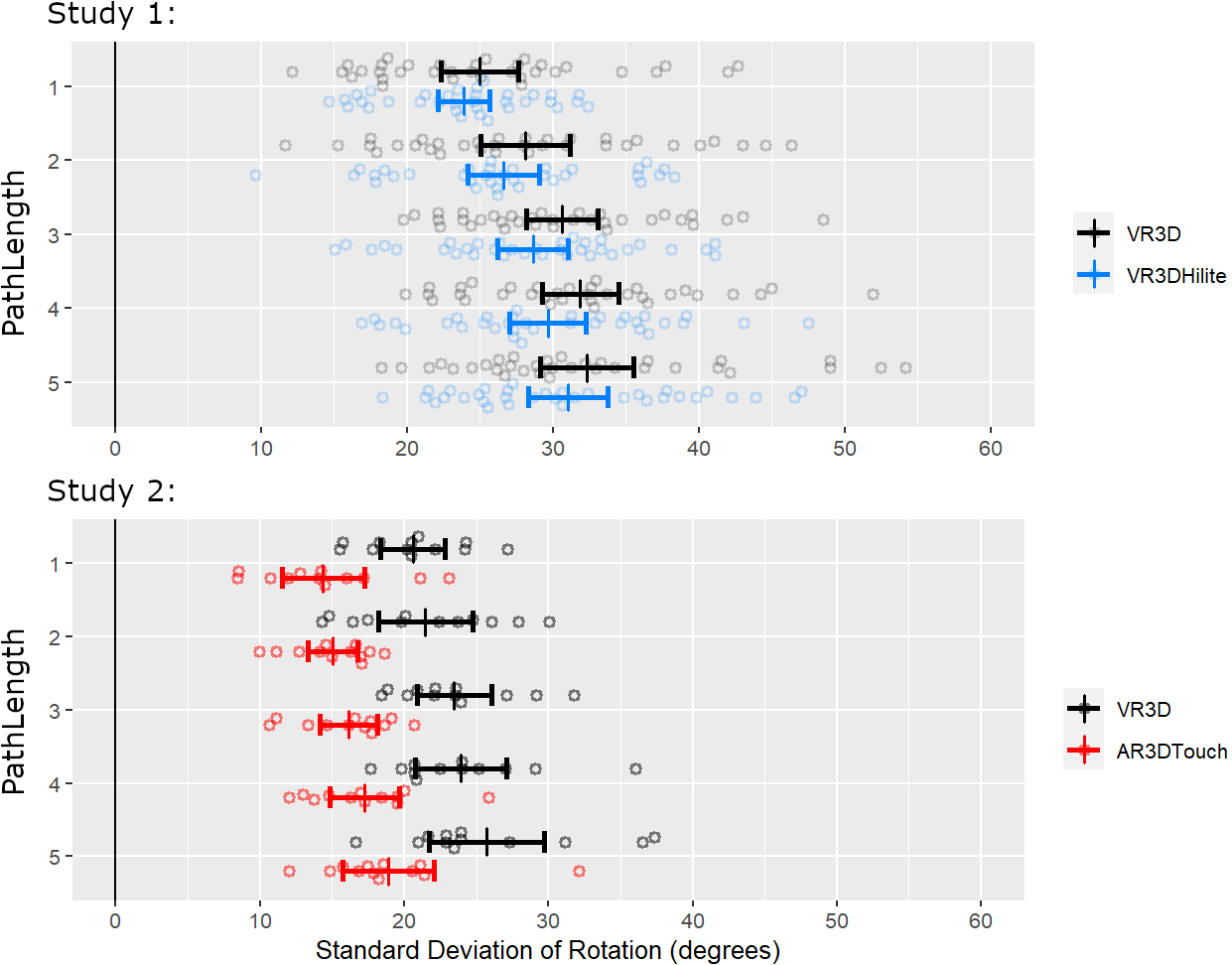}
 \caption{
    How much users rotated their view of the network (Section~\ref{sec:rotationCalculation}).
  }
 \label{fig:results-rotation}
\end{figure}

\begin{figure}[tb]
 \centering
 \includegraphics[width=\columnwidth]{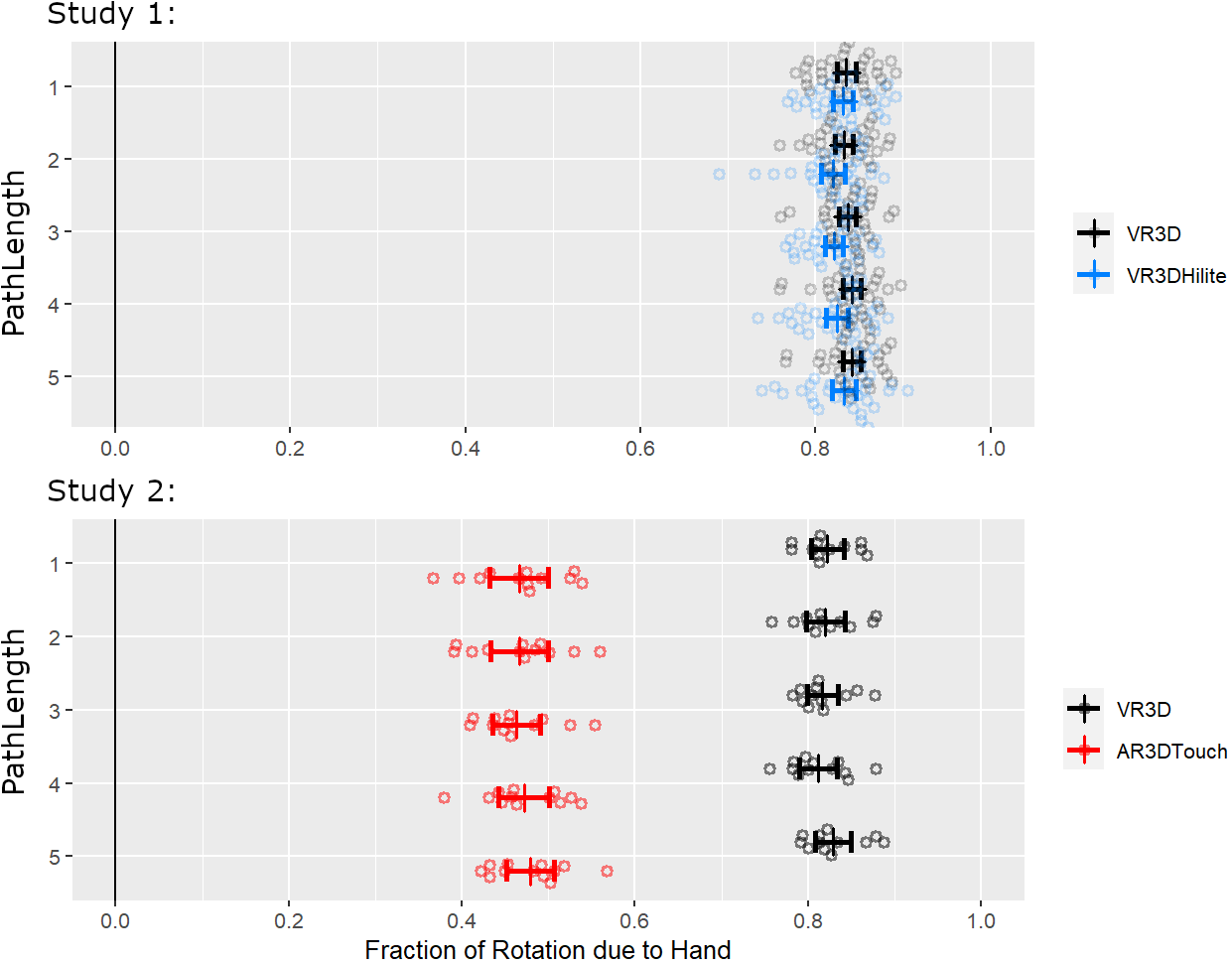}
 \caption{
    How much of the rotation was due to hand motion,
    as opposed to head motion
    (Section~\ref{sec:rotationCalculation}).
  }
 \label{fig:results-rotationFraction}
\end{figure}

As mentioned earlier (Section~\ref{sec:study1predictions}), in the
W+M study, ``the mean Euclidean distance between start nodes and end nodes was the same'' \cite{ware2008}.
Because our paths could be up to 5 edges long,
this kind of control was not feasible in our experiment,
nor would it have yielded realistic tasks.
However, to check how much the Euclidean distance
may have influenced user responses in our Study 1,
we computed two additional variables for each trial:
first, $\Delta$ = response $-$ \PathLength,
so that $\Delta$ is negative, zero, or positive when the user's response
is under, equal to, or over the correct \PathLength, respectively.
For example, if the user's response is 5 when the \PathLength\ is actually
3, then $\Delta = 2$.
Second, we found all shortest paths in the network and,
for each \PathLength,
we found the average and standard deviation of the
Euclidean distances (from end-node to end-node) of those paths.
This allowed us to compute a $z$-score for any pair of nodes,
as a way to
compare their Euclidean distance to that of other pairs of nodes
with the same topological distance.
So in a given trial, if the shortest path between the end-nodes
is 3 edges, but the $z$-score for those end-nodes is greater than 1,
this means that shortest paths of 3 edges in that network tend to have
end-nodes that are closer (in the Euclidean sense),
and we might expect such a $z$-score to bias the user
toward over-estimating the \PathLength\ in that trial.
To test for such a bias, we checked for a correlation between $\Delta$ and the $z$-score.
The correlation test yielded $R < 0.005$ and $p > 0.05$,
hence no evidence of the Euclidean distance biasing the user toward erroneous responses.

During the conditions without mouse,
where the DH was free,
3 out of the 34 users (plus 1 other user from the pilot) were either observed lifting their DH toward the controller during trials, as if trying to touch
the virtual network, and/or described imagining touching or
wanting to touch the virtual network during the post-questionnaire discussion.  One of these users said this may have been because of their previous experience with an Oculus Quest 2 which displays the user's hands, and another user suggested using AR to allow the user to see their own hands.

\subsection{Discussion} %

As discussed in Section~\ref{sec:study1predictions},
the lower error rate in 3D may be due in part to the shortest paths
being more straight in 3D, because the layout algorithm has more freedom to position nodes.
For each of the shortest paths in the trials of Study 1,
we computed the following ratio: the sum of the lengths of the edges in the path divided by the Euclidean distance
between the two end-nodes of the path.
This ratio is high if the path is circuitous (i.e., winding), but close to 1 if the path is straight.
The average ratio for our 2D trials was 1.509, but for 3D it was 1.400, thus more straight.
The difficulty of a more winding path may be related to studies finding that participants
take longer to trace curves connecting two targets when the curve is longer,
even when the Euclidean distance between the targets is the same \cite{crundall2008}.

Figure~\ref{fig:results-rotationFraction} suggests
that most of the benefit of rotation (i.e., of motion parallax) comes from
motion of the hand rather than of the head.
Figures 6 and 7 in \cite{ware2008} suggest that motion provides
at least as much benefit as stereo.
Taken together, these suggest that a user would benefit simply from the ability to rotate
a 3D visualization with their hand,
without any headset, stereo, or head-coupled perspective.

We observed that some users wished they could touch the networks.
Previous work \cite{drogemuller2021} found that users preferred physical networks that could be touched.
This motivates our next section.

\section{Study 2: 3D Virtual and Physical}

Study 2 compared virtual and physical representations of networks,
using a VR and an AR headset, respectively,
and the same task as Study 1.
Study 2 was more exploratory,
where the number of participants
was determined by convenience,
and not preregistered.

\subsection{Pilot and Choice of Main Conditions} %

\begin{figure}[tb]
 \centering
 \includegraphics[width=\columnwidth]{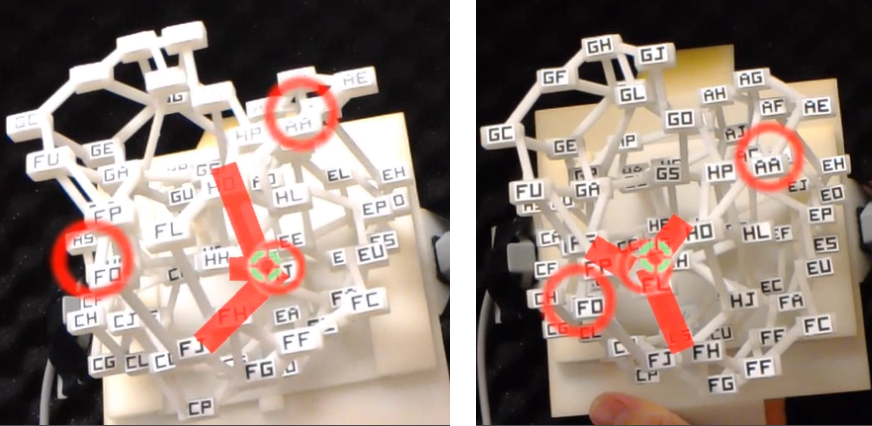}
 \caption{
    Prototype implementation of highlighting of physical edges.  Nodes FO and AA are indicated with red circular rings. The mouse cursor (green) hovers over HJ (Left) and FL (Right), causing three incident edges to highlight in each case.
    This was not used in our studies due to insufficient
    tracking accuracy.
  }
 \label{fig:physicalEdgeHiliting}
\end{figure}

To display virtual information on top of a 3D printed physical network,
which we call {\bf augmented physicalization},
we need some way for the AR headset to know where the network is located.
We tested 3 different ways of tracking the physical network,
including using the headset's built-in camera and a combination of 3 external cameras.
Section~\ref{sec:study2hardware} describes our ultimate tracking method.
Figure~\ref{fig:physicalEdgeHiliting} shows a prototype \ARHilite\ condition,
where the user's DH moves a mouse, and the AR headset displays a virtual cursor and virtual
highlighting on parts of the network.
Unfortunately, we were unable to achieve the accuracy
necessary to clearly highlight individual nodes or edges on a physical network.
We suspect that part of the problem is due to small errors in the IPD
used to generate the stereo rendering on the headset, making virtual imagery slightly
misaligned with physical objects.
We thus dropped the \ARHilite\ condition and do not display the circular rings
around nodes in any conditions of Study 2.
Nevertheless, the end-nodes in Study 2 are indicated with callout line segments.
In VR, these callouts are precisely located, but in AR, there are errors of $\approx$2-3cm
in their apparent locations.
Despite having only approximately
correct positions in AR,
the callouts do help the user find the correct physical end-nodes faster than if the user had no visual aid,
and constitute an example of augmented physicalization.
We expect that the approximate locations in AR
will slow down users compared to VR,
but it is plausible that this will have
minimal impact on the user's error rate in AR.

\begin{figure}[tb]
 \centering
 \includegraphics[width=\columnwidth]{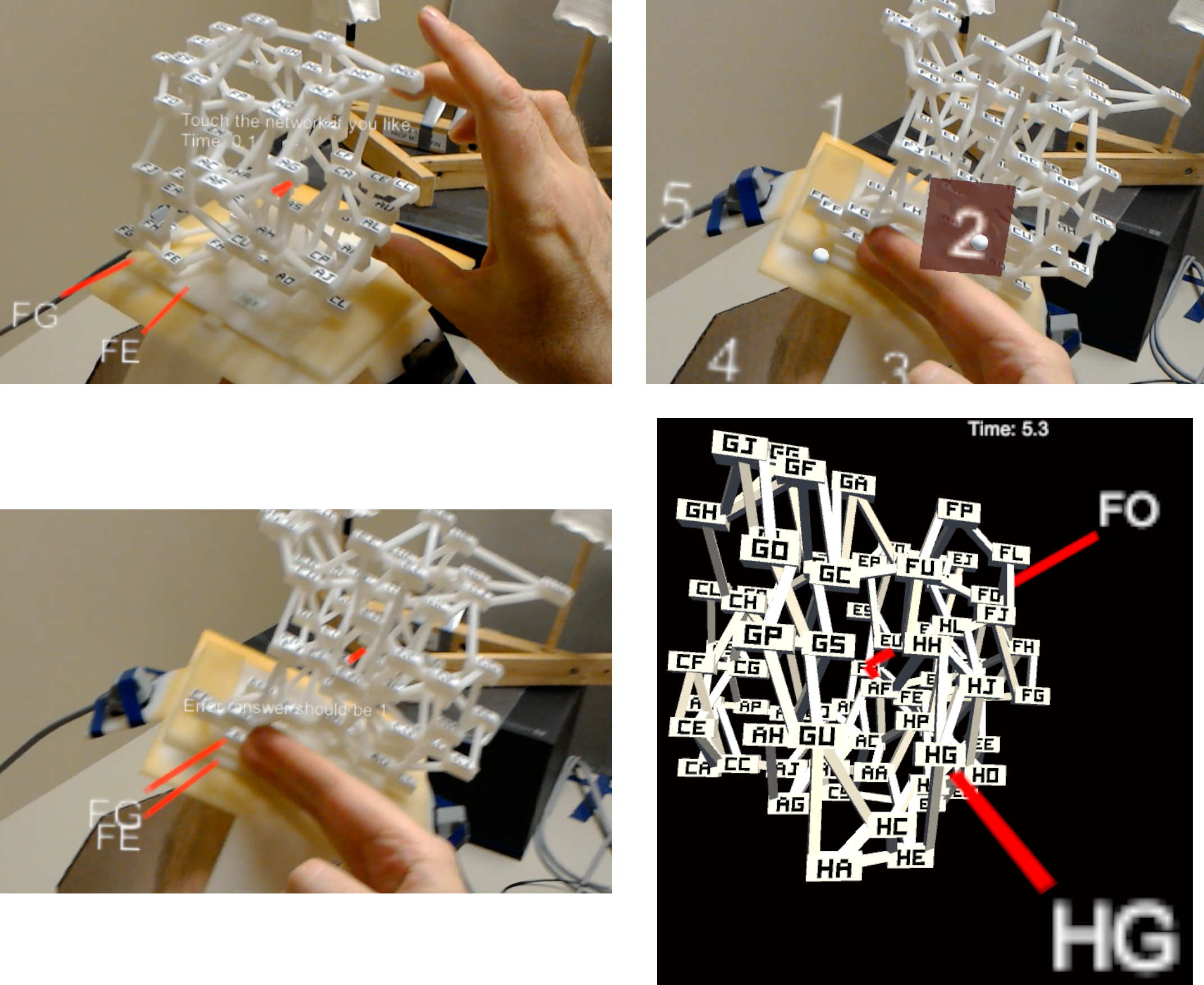}
 \caption{
    The first three images show a trial in the \ARTouch\ condition of study 2.
    The user touches the indicated nodes
    FG and FE and answers ``2'' (Top Right)
    only to be shown the error feedback
    ``Error, answer should be 1''
    (Bottom Left).
    The last image (Bottom Right) shows the variant of the
    \VRThree\ condition that was used in study 2.
    In both conditions, nodes are indicated with red callout line segments but not
    circular rings, due to insufficient tracking accuracy in AR.
  }
 \label{fig:study2}
\end{figure}

We ran a pilot with 3 users
and 4 conditions: \VRThree, \VRThreeHilite, \AR, and \ARTouch.
In the \AR\ condition,
the user repositions the network with their NDH but may not touch the network with their DH.
In \ARTouch\ (Figure~\ref{fig:study2}), the user is encouraged to touch the network with their DH.
For each of the 4 conditions,
the user performed 10 warmup trials,
followed by 20 trials with each of two networks.
The ordering of headsets was random, as was the ordering of the pair of main conditions
within each headset.
In contrast with Study 1,
each user was given more explanation of how the mouse worked in 3D for the \VRThreeHilite\ condition.
Sessions lasted 2 hours per user.
All users chose \AR\ as their least favorite condition;
one user found it uncomfortable to wear the HoloLens for so long;
and two users reported that the tracking accuracy
got worse over time (this was probably due to the users
holding the network in different positions during
calibration and during trials).
Therefore, some changes were made for the final Study 2 experiment:
we eliminated the \AR\ condition, which was the least favorite condition of the users
and less realistic than \ARTouch;
we also eliminated the \VRThreeHilite\ condition,
because most users in Study 1 did not find the mouse in \VRThreeHilite\ useful,
and we wanted Study 2 to involve the same number of trials with each headset.
We also increased the number of opportunities for the user to take breaks in both conditions,
and opportunities to redo the calibration
during the \ARTouch\ condition.
With only 2 main conditions,
we could also slightly increase the number of trials per condition while also
decreasing the total duration of each user session (Section~\ref{sec:study2design}).

Thus, Study 2 had two values for \MainCondition:
\VRThree, and \ARTouch.

As detailed in the next section, the AR condition suffered from a smaller FOV,
latency, and tracking error,
compared with the VR condition.
This creates confounds,
however whichever condition outperforms the other,
the results could be informative:
if \ARTouch\ yields a lower error rate,
despite the shortcomings of the AR system,
this demonstrates the importance of physical realism and/or
the ability to touch the network;
and if \VRThree\ yields a lower error rate,
despite affording no way
to interact with the network,
this shows how important it is for AR systems to be improved
to reach their full potential.

\subsection{Hardware} \label{sec:study2hardware} %

For the \VRThree\ condition, the same headset and controller were used
as in the previous study.
For \ARTouch,
we used a Microsoft HoloLens headset,
which has 1268$\times$720 pixels per eye. %
The FOV for displaying virtual information is limited to $\approx$31$\times$17$^\circ$,
however the physical world is visible through a much wider FOV.
The virtual images are rendered at a fixed focal distance
of $\approx$2 meters\cite{hololensFocalDistance}
however stereo and vergence depth cues create the illusion of virtual imagery at any distance.
We measured a framerate of 30 fps.

\begin{figure}[tb]
 \centering
 \includegraphics[width=\columnwidth]{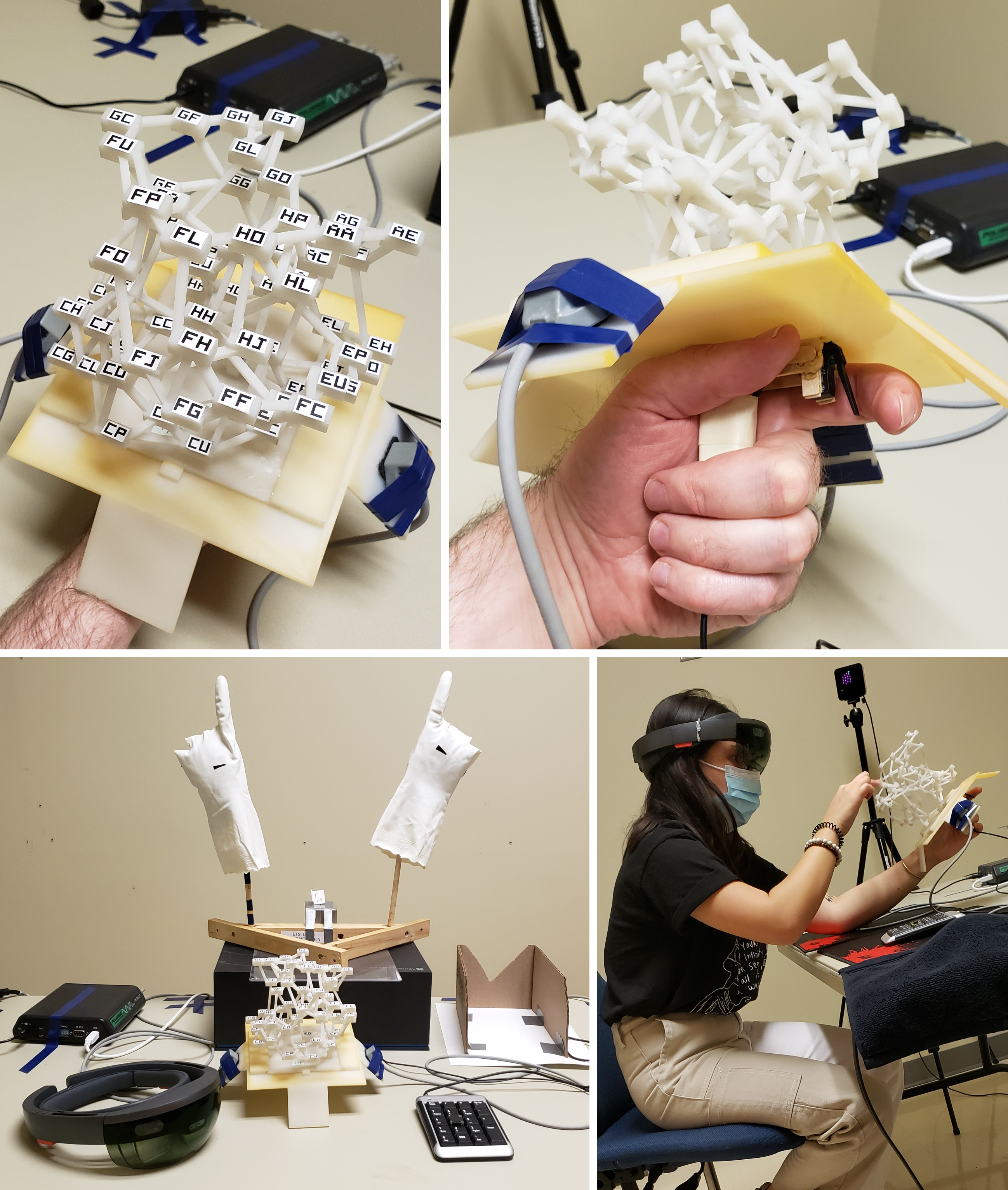}
 \caption{
    The equipment for the \ARTouch\ condition of Study 2.
    Top: the network holder, held in the NDH,
    with a trigger button.
    Bottom left: 
    registration rig with 2 fake hands,
    Microsoft HoloLens headset,
    network holder,
    and keypad.
    Bottom right: during a trial.
  }
 \label{fig:study2hardware}
\end{figure}

3D printed networks were held in a ``network holder'' by the user's NDH.
A Polhemus Patriot reported the positions and orientations (in the Patriot's local coordinate system) of two sensors
attached to the network holder.
Because the HoloLens has built-in functionality to detect hands and report their 3D position in the headset's coordinate system,
we constructed a rig with two fake hands (Figure~\ref{fig:study2hardware}(Bottom Left)) whose positions were fixed with respect to the Patriot,
allowing us to determine the position of the network with respect to the headset with an accuracy of $\approx$5cm.
The user also performed a simple calibration procedure
to further improve the accuracy of the tracking to $\approx$2-3cm.

The Patriot was connected to the same PC mentioned in Section~\ref{sec:study1hardware}.
This PC processed the position and orientation information
from the Patriot and transmitted it via UDP packets over wifi to the HoloLens.  Although the Patriot can read information at 60 Hz,
we only transmitted 10 packets per second to the HoloLens,
because a higher rate led to dropped packets.
By studying footage recorded through the HoloLens, we estimate that the latency between moving the network holder
and the HoloLens updating its rendered virtual imagery was $\approx$150-250ms,
despite using a dedicated high bandwidth wifi router (ASUS RT-AC5300). %

Having two different headsets in Study 2, with differences in
FOV and other characteristics,
necessarily introduces confounds.
An alternative approach would have been to use the HoloLens in both conditions of Study 2,
however this would have meant imposing a limited FOV that is not representative of the state-of-the-art in VR.
Although the differences in headsets will certainly create differences
in the time taken for trials,
we are more interested in differences in error rate
between the main conditions,
and it is plausible that error rate will be less affected
by the differences in the headsets.

\subsection{Use of blur in virtual feedback}

Like most headsets,
our VR and AR headsets
each reproduce correct stereo disparity
and vergence depth cues,
but not correct accommodation depth cues,
due to the virtual imagery being rendered at a fixed focal distance.
In VR, this results in the well known ``vergence-accommodation conflict'', which is often barely noticeable.
However, with see-through AR headsets like the HoloLens,
there is a further challenge:
if virtual imagery (such as a virtual highlight) is rendered at the same location as
a physical object (such as part of a physical network),
it is impossible for the user to accommodate (i.e., ``focus on'')
both simultaneously.  Although both may appear to be 30cm from
the user's eyes in terms of stereo disparity and vergence depth cues, the focal distance of all virtual imagery rendered by the HoloLens is $\approx$2 meters\cite{hololensFocalDistance}.

To avoid having users focus on such virtual feedback,
making the physical network appear blurry,
we render callouts with a blur
effect (Figure~\ref{fig:study2}), i.e., without sharp edges.
For consistency, this was done in both \VRThree\ and \ARTouch.

\subsection{Users} %

12 new users were recruited:
9 male, 3 female;
all right handed;
age 20 to 42 years (average 27.3);
IPD 57 to 70mm (average 62.7);
stereoacuity test scores 3/10 to 10/10 (average 8.9).
As in the previous study (Section~\ref{sec:study1protocol}),
users were shown several printouts of example trials,
with conditions in random order,
to explain the task,
and each headset was adjusted to the user's IPD
before beginning warmup trials.

\subsection{Design}\label{sec:study2design} %

Each user experienced the
2 levels of \MainCondition\ \{\VRThree, \ARTouch\} in random order.
For each \MainCondition,
the user performed 10 warmup trials in random order
(5 levels of \PathLength\ $\times$ 2 repetitions)
with the warmup network,
followed by 15 trials in random order
(5 levels of \PathLength\ $\times$ 3 repetitions)
with another network chosen at random,
followed by another 5 warmup trials with the warmup network,
15 trials with another network,
5 warmup trials with the warmup network,
and another 15 trials with another network.
Each subsequence of warmup trials gave the user an opportunity to
take a break, adjust the headset, and redo the calibration if they wished.
There were a total of
12 users
$\times$ 2 levels of \MainCondition\ 
$\times$ 3 networks
$\times$ 5 levels of \PathLength\
$\times$ 3 repetitions per trial
= 1080 trials, not counting warmup trials and not counting pilot data.
Each session with a user lasted $\approx$ 1.5 hours.

\subsection{Results} \label{sec:study2results} %

\begin{figure}[tb]
 \centering
 \includegraphics[width=\columnwidth]{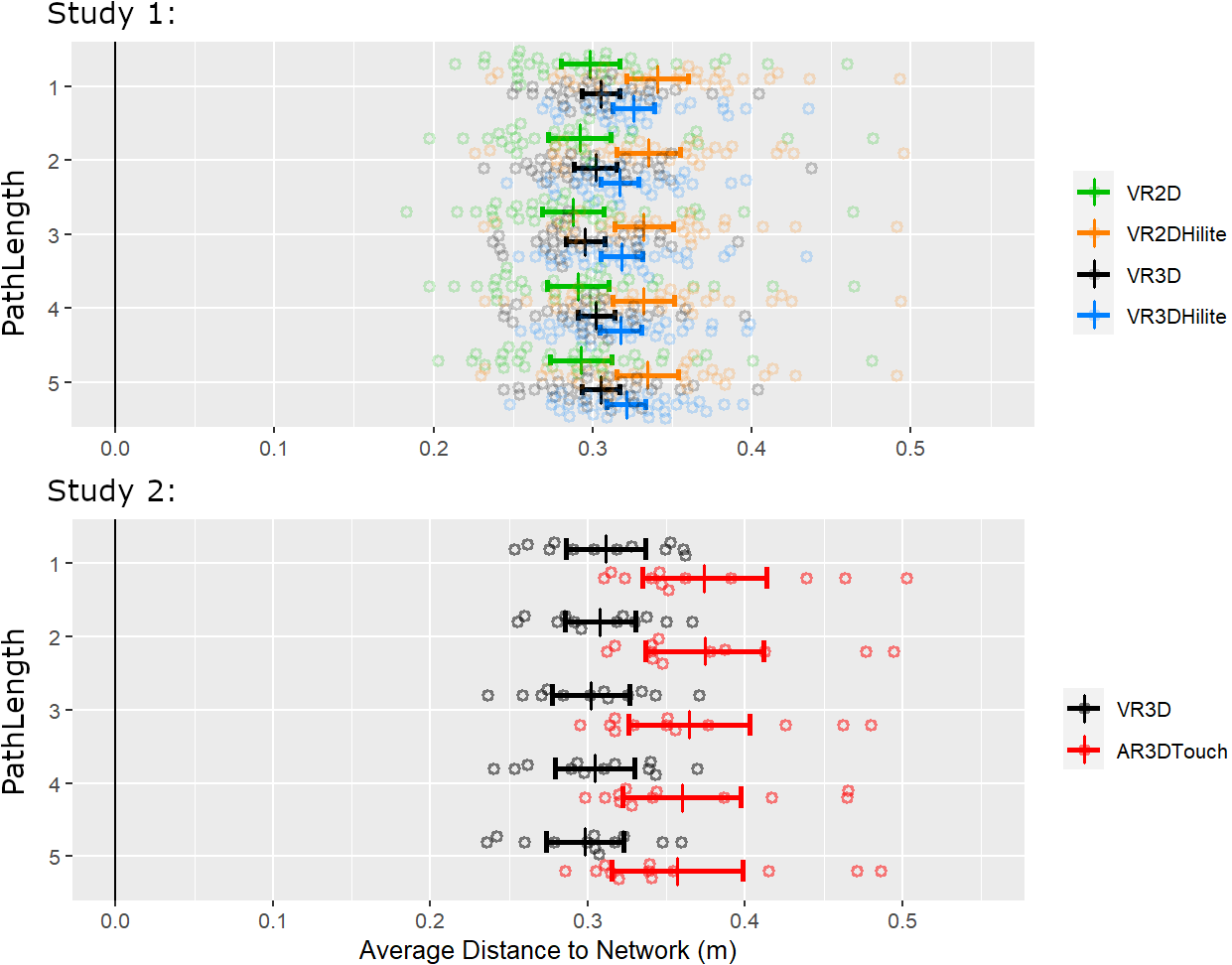}
 \caption{
    How far users held the network from their head.
  }
 \label{fig:results-distance}
\end{figure}

We find no evidence of a difference in error rate
between the \ARTouch\ and \VRThree\ conditions
(Figure~\ref{fig:results-errorRate}).
In \ARTouch, users took more time
(Figure~\ref{fig:results-time}),
which is explained by the difficulty that users had
in finding the end-nodes at the start of each trial.
Users also
rotated their view less
(Figure~\ref{fig:results-rotation})
and rotated less with their hand and more with head motion
(Figure~\ref{fig:results-rotationFraction}),
which is explained by user feedback
indicating that the network holder in \ARTouch\ was
not as easy to rotate as the handheld controller
in the \VRThree\ condition.
Users also held the network further away from their
head
(Figure~\ref{fig:results-distance}),
which is explained by the more limited FOV in AR.

All 12 users preferred the \VRThree\ condition
to \ARTouch\ (Figure~\ref{fig:subjective}).
The reasons given by users for this preference were:
the inaccurate positioning of callouts in AR (mentioned by 9 out of the 12 users);
the limited FOV of the AR headset (mentioned by 4 users);
the AR headset being less comfortable (4);
the network holder in the \ARTouch\ condition being more difficult to rotate than the VR controller (4);
the \VRThree\ condition providing better visual contrast
between the network and the black background (4);
and the latency with the AR system (1).

Based on user feedback, users often performed
each trial of \ARTouch\ in two stages: first, identifying the end-nodes,
and second, finding the shortest path between them.
The first was difficult because of the limited FOV and the
positional error in the callouts.
Once the end-nodes were identified, users would often ``mark'' them
by touching their fingertips to one or both end-nodes
and maintaining contact while
looking for the shortest path.
Several other behaviors were observed with users' fingers:
touching intermediate nodes along a path (observed in 9 out of the 12 users),
touching multiple nodes simultaneously (8 users),
pivoting the network around the hand while maintaining contact with one or more DH fingers (7),
pointing at nodes without touching them (6),
and grabbing a node with 2 fingers (3).

We also noted which fingers of the DH were employed during the \ARTouch\ condition.
Letters t, i, m, r, p denote thumb, index, middle, ring,
and pinky fingers, respectively,
and multiple letters indicate combinations,
such as ti for thumb + index.
Fingers i, m, p were often employed individually.
We also observed simultaneous uses
of fingers:
im (used by 7 out of the 12 users),
ti (6 users),
tm (5),
ip (2),
ir (1),
mp (1),
tim (1),
imr (1).

Other notable behaviors observed were:
touching edges in addition to nodes;
tapping a sequence of nodes along a path;
sliding a fingertip along the edges of a path;
walking two fingers (index and middle) along the nodes of a path like legs;
touching one node and pivoting the hand around the network, while maintaining contact with the finger;
touching three nodes at once (either with tim or with imr).

One user %
explained that they used 2 fingers simultaneously to trace 2 different paths
between end-nodes.
Another  %
said they would have preferred the \ARTouch\ condition over \VRThree\ if the problems with FOV and tracking accuracy were fixed.
Another user %
said that touch had value for tracking the path with the finger.
Another %
stated that they really wanted to touch the network in the \VRThree\ condition, and wanted to see their fingers in VR, and that they were running their fingers through space where the path would have been during the \VRThree\ condition.
Another user (during the pilot)  %
actually preferred the \ARTouch\ condition to \VRThree\ despite the limitations of the AR system, stating that the physical network allowed the task to be done faster and more simply.

\subsection{Discussion} \label{sec:study2discussion} %

In the \ARTouch\ condition, users employed their fingers
in a variety of manners.
The different uses of hands
has been studied before in visualizations \cite{walny2017}
and physicalizations \cite{taher2017emerge,drogemuller2021}.
Other work \cite{stival2019} %
has proposed a taxonomy
of uses of hands for grasping.
We are unaware of a taxonomy of more general uses of hands
relevant to data physicalizations.

Users preferred \VRThree\ over \ARTouch, but the reasons
given for this are almost entirely due to technological limitations of the AR platform.
Physicalizations that can be touched have been shown to
be preferred \cite{drogemuller2021} or yielded better performance \cite{jansen2013evaluating}
than 3D visualizations.
Hence, next steps could be to either improve the AR platform,
or improve the VR platform to better support the way users leverage their fingers.
We now consider each of these.

Our AR platform suffers from inaccurate alignment of virtual and physical objects.
Before using the Polhemus Patriot for tracking, we tried
using the RGB camera on the HoloLens
as well as multiple external cameras for object tracking,
but none of these methods achieved acceptable accuracy.
An alternative approach would be to use
video pass-through AR (either with a headset or not,
as in \cite{gillet2005}) to avoid the need for highly accurate
tracking, and enable virtual highlighting of physical
nodes and edges (Figure~\ref{fig:physicalEdgeHiliting})
with pixel-precise alignment,
correct occlusion cues,
and no conflicting accommodation (focal) distances.

To improve the VR platform to better support finger interaction,
we notice that users in our \ARTouch\ condition primarily
used fingers to mark parts of the network.
For example, users would often touch one or more fingertips against parts of the network, and then pivot a hand
(to examine the network from a different view)
while {\em maintaining contact} with fingertips, which was {\em made easier by friction}.
This suggests an interaction technique where the user can use their DH to define one or more ``sticky marks'' on the network,
that remain even while the hands continue to move.
We also observed users sliding their fingers along edges,
    which suggests support for ``slippery marks''
    whose positions are updated to slide along edges,
    maybe as if being pulled by strings attached to the DH.
Although physical fingers are limited to maintaining one mark per finger,
    this needn't be the case with virtual marks: different fingers might be pinched against the thumb to
    invoke different commands to create, modify, or delete each of many marks. %
For example, when the user's DH approaches a virtual network,
a rope cursor \cite{guillon2015rope} from the DH's thumb could extend to the nearest node,
and a DH pinch against the thumb by the index, middle, or ring finger could create a sticky mark, create a slippery mark, or delete the mark, respectively, at that node.
Such finger-based interaction might be generally useful across many tasks
beyond finding shortest paths.

\section{Conclusions}

Our Study 1 provides strong evidence that path tracing
is less error prone in 3D layouts than in 2D layouts
(Figure~\ref{fig:results-study1-prediction1}),
despite the use of edge-routing in 2D.
The use of mouse-driven interactive highlighting in 2D
reduces the error rate in 2D, but not as much as
using a 3D layout (Figure~\ref{fig:results-study1-errorRate2}).
3D was also the most preferred layout (Figure~\ref{fig:subjective})

Our Study 2 found no evidence of different error rates
between the virtual (VR) and physical (AR) conditions
(Figure~\ref{fig:results-errorRate}).
VR was preferred by users,
but this was largely due to technological
limitations of the AR platform,
and users touched the physicalized networks in a variety of ways.

\section{Future Directions}

Section~\ref{sec:study2discussion} described ideas for future work.
In addition, to benefit from the advantages of 2D and 3D,
techniques could be designed allowing a user to rapidly
switch between 2D and 3D layouts,
possibly mixing styles in a hybrid that is
2D for most of the network
but 3D in certain parts where the user is more interested in perceiving shortest
paths.
New fabrication methods might also enable physicalized networks
that contain embedded buttons, touch sensors, and/or lights,
for richer interaction.

\ifthenelse{\boolean{IncludeAppendix}}{

\appendix[Calibration of HoloLens]

We implemented calibration with a homography
matrix that transforms from the Patriot's 3D space
to the 3D coordinate system of the HoloLens.
Ideally, the calibration would have been done only once by a
researcher, and the resulting homography matrix saved,
so that users wouldn't need to perform any calibration.
However, there are nonnegligible errors in the positions
reported by
the Patriot and the HoloLens,
and the best homography to use changes from one session to the next.
A further difficulty arises from the
coordinate system of the HoloLens being different
for each session: its origin and axes are defined
by however the user's head happens to be positioned and
oriented when our client code begins executing.
We now describe how we kept the calibration as simple
as possible for users, while estimating a good quality homography.

We
extended the 2D-to-2D homography calculation given in \cite{kriegman2007}, and
implemented a subroutine that takes as input
a list of points whose 3D coordinates are known in
two different coordinate systems, and outputs a
4$\times$4 homography matrix to transform between the two
coordinate systems.
Because the HoloLens reports the positions of the fake hands
in its own coordinate system, and because we know (roughly)
where the hands are located in the Patriot's coordinate system,
we can use our subroutine to compute
an initial homography $h_0$.
It is possible, however, to obtain a better homography $H_0$
by having the user go through a 9-point calibration procedure that we implemented:
in this procedure,
the HoloLens displays a sequence of 9 virtual points
(located at the corners and center of a cube occupying the
user's working volume), and the user holds a Patriot sensor
at each of the displayed locations, allowing our client code to
obtain the coordinates of each of the 9 points in both coordinate
systems, and then use the same subroutine to compute
the improved $H_0$.
However, we did not want each participant in our experiment
to have to perform this 9-point calibration,
as it is time-consuming to explain and perform carefully.

\begin{figure}[tb]
 \centering
 \includegraphics[width=\columnwidth]{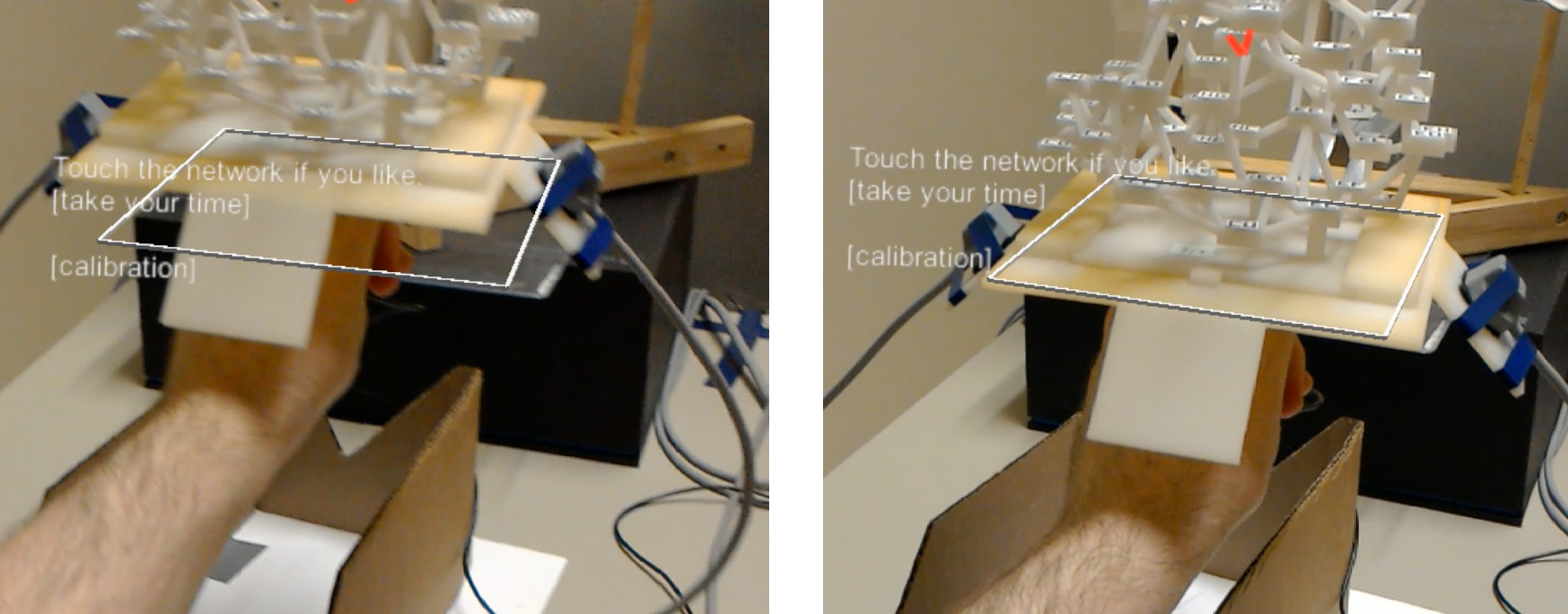}
 \caption{
    In the \ARTouch\ condition of study 2,
    the user performs calibration by moving the network holder
    onto a white virtual square.
  }
 \label{fig:calibration}
\end{figure}

We speculated
that each homography $h = v f$
can be factored into two components,
a variable component $v$ that changes from one session to the next
depending on noise and on how the headset initializes its coordinate system, and a fixed component $f$ that depends on the placement
of the fake hands with respect to the Patriot.
We won't try to extract these components,
but there is an indirect way of improving the estimate of $f$
and therefore improving $h$.
Consider the $h_0$ and $H_0$ defined in the previous paragraph,
where the former is a lower quality homography, and the latter
is a higher quality one obtained with 9-point calibration.
Each has factors $h_0 = v_0 f$ and $H_0 = v_0 F$,
where $F$ is a better estimate of the fixed parameters
that don't change from one session to the next.
Compute and save $Q = h_0^{-1} H_0$.
In any future session $i$,
compute $h_i$ (based on the positions of the fake hands)
and then multiply by the previously saved $Q$
to obtain $H_i = h_i Q$ which is a better homography than
$h_i$.
To see why,
notice that
$H_i = h_i Q = h_i h_0^{-1} H_0 = h_i (v_0 f)^{-1} v_0 F
= v_i f (f^{-1} v_0^{-1}) v_0 F = v_i F$.
In other words, the multiplication by $Q$
effectively replaces the lower quality $f$
with the higher quality $F$.
The 9-point procedure is only performed once by a researcher to obtain $Q$
and needn't be repeated by each participant.
We implemented this and found that it noticeably improved the
initial tracking accuracy.
However, to improve accuracy even more,
each participant
also performed the simple calibration procedure
shown in Figure~\ref{fig:calibration}.

}{} %

\ifCLASSOPTIONcompsoc
  \section*{Acknowledgments}
\else
  \section*{Acknowledgment}
\fi

Thanks
to Dylan R. McGuffin for mounting the trigger button
and building the registration rig;
to Alicia Servera for help building an earlier rig;
to Steve Haroz and Pierre Dragicevic for detailed
advice on experiment design and statistical analysis;
to Nathalie Henry Riche for the idea of comparing with physical networks;
to Eyal Ofek, Ken Hinckley, and Jean-Daniel Fekete for general discussion;
to Lev Nachmanson and Tim Dwyer for advice
on edge-routing;
to Fran\c{c}ois B\'{e}rard for ideas on measuring latency;
to Mar Gonzalez Franco and Bill Buxton for pointers to related work;
to Peter Vale and Erwan Normand for programming tips;
and to Minda Miloff for coaching.
This research was supported by
Microsoft Research and NSERC.

\ifCLASSOPTIONcaptionsoff
  \newpage
\fi

\bibliography{main}
\bibliographystyle{IEEEtran}

\begin{IEEEbiography}[{\includegraphics[width=1in]{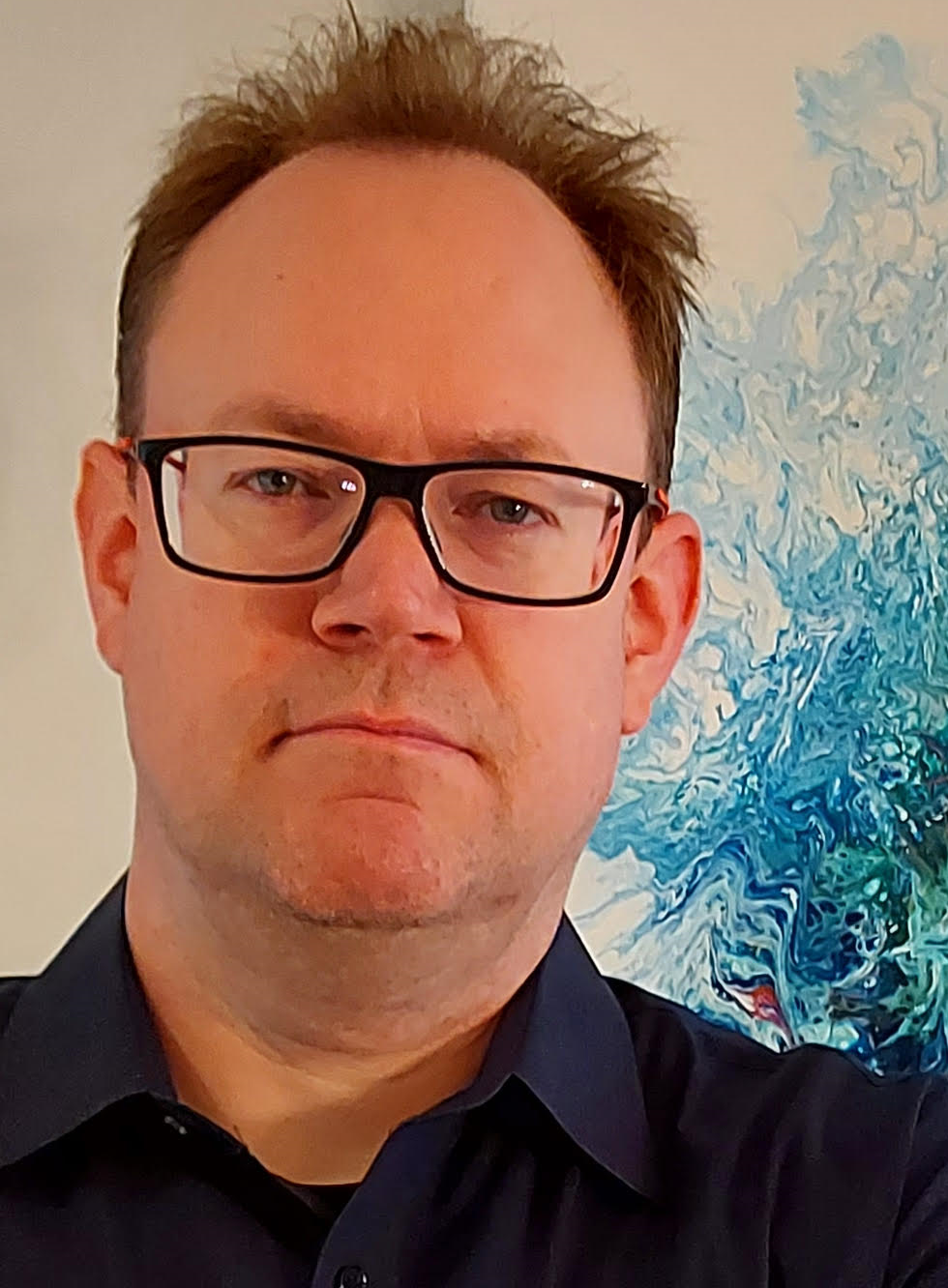}}]{Michael J. McGuffin} is a full professor at ETS, a French-language engineering school in Montreal, Canada, where his students do research in HCI and visualization. His recent interests include visual programming
and artificial life.
In 2009, his paper at the IEEE Information Visualization Conference (InfoVis 2009) received an Honorable Mention.
\end{IEEEbiography}

\begin{IEEEbiography}[{\includegraphics[width=1in]{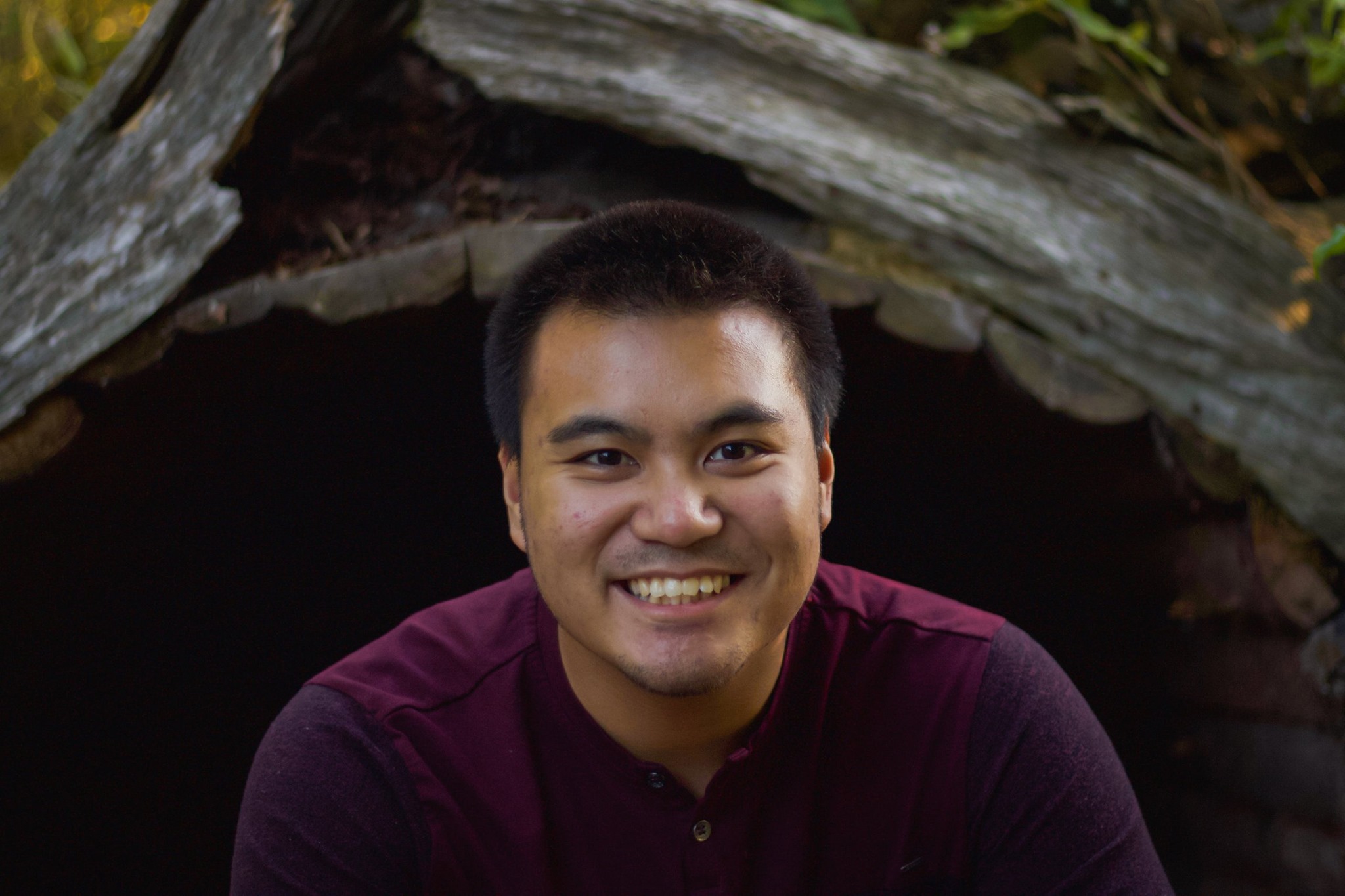}}]{Ryan Servera} is a software developer. He received a Bachelor's of engineering in electrical engineering with a minor in software engineering at McGill University in 2020. In 2019, he interned as a research assistant at ETS where he was able to explore VR and AR. He is particularly interested in human computer interaction in video games.
\end{IEEEbiography}

\begin{IEEEbiography}[{\includegraphics[width=1in]{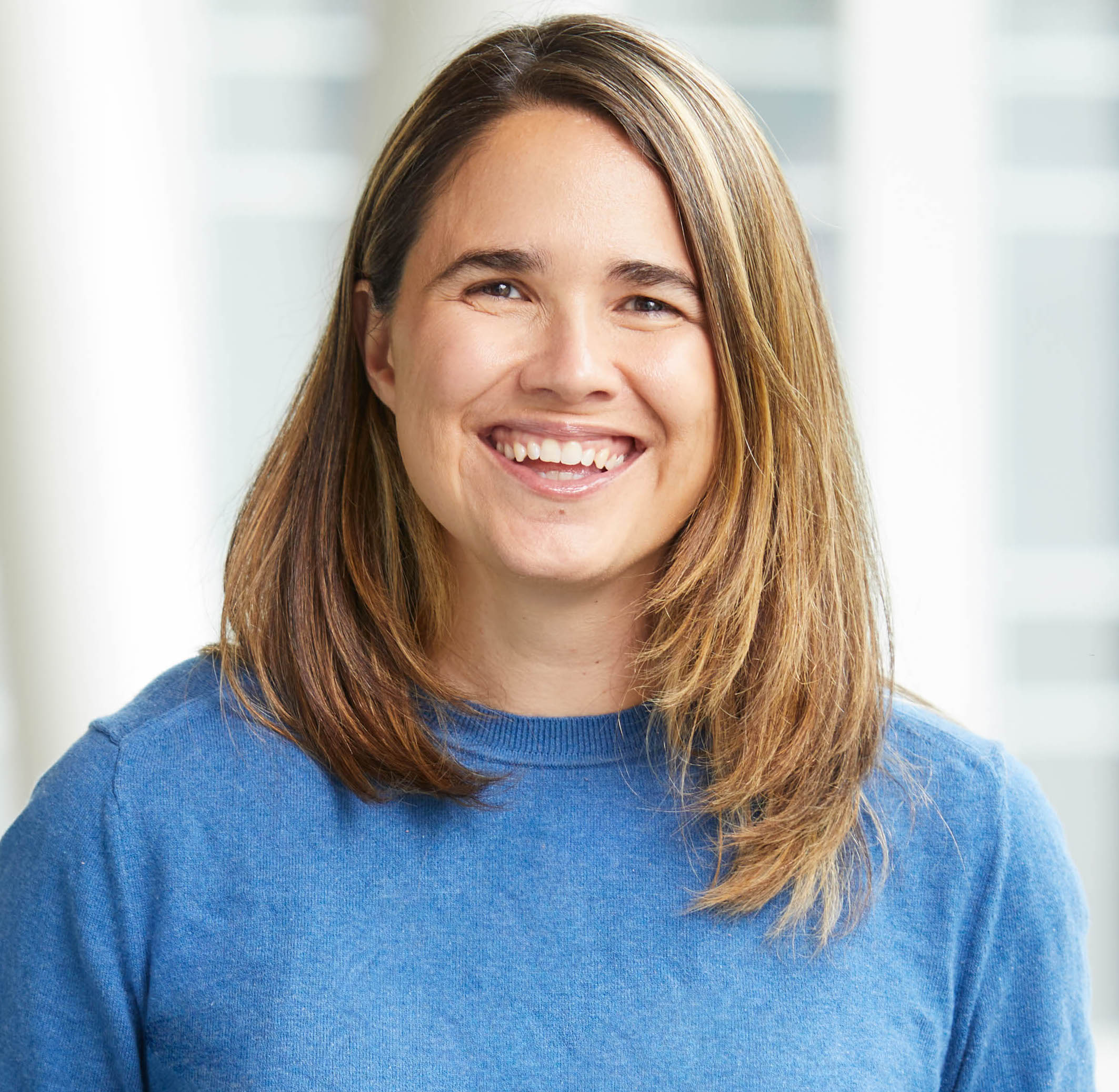}}]{Marie Forest} is a Lecturer in Statistics at ETS, with a DPhil in Statistics from the University of Oxford. She is interested in the development of web application teaching tools to help students understand probability and statistics through simulations. In the past, her research areas were in the analysis of genomic data and stochastic modelling applied to genetics.
\end{IEEEbiography}

\end{document}

\begin{center}
\noindent {\bf Path Tracing in 2D, 3D, and Physicalized Networks: Supplemental} \\
~\\
Michael J. McGuffin, Ryan Servera, Marie Forest
\end{center}

\newcommand{\mySection}[1]{
\noindent {\bf #1}
}

\mySection{Hardware}

We measured the masses of the devices used:\\
HTC Vive headset: $\approx$615g\\ %
One HTC Vive handheld controller: $\approx$200g\\
HoloLens headset: $\approx$577g\\
3D printed network + network holder: $\approx$370-378g, depending on the network\\

\mySection{Additional Results from Study 1}

We measured the hand span
(distance between tips of thumb and pinky with fingers spread out) of each user.
We also computed the overall error rate of each user
in 3D (computed from trials in 3D, not including warmups, and only including path lengths of 2, 3, 4, because path lengths of 1 and 5 yielded lower error rates that are less interesting),
and the overall amount of rotation performed by users
in 3D (not including warmups).

We found little to no evidence of a correlation between
IPD and error rate,
IPD and stereo acuity,
age and error rate,
hand span and rotation,
hand span and error rate,
stereo acuity and error rate.
We also found no evidence of a difference in the error rates
between male and female participants. \\

\mySection{Rendering on the HoloLens}

``Focus point'' hinting was used to stabilize the image rendered by the
HoloLens\\
( https://docs.microsoft.com/en-us/windows/mixed-reality/develop/unity/focus-point-in-unity ),
and rendering of polygons was done with a flat shader
to optimize the framerate. \\

\mySection{Tracking of Networks in Study 2}

Before deciding to use the Polhemus Patriot for tracking,
we previously implemented and tested two alternative tracking methods,
first using the RGB camera on the HoloLens with Vuforia,
and second using three simultaneous external cameras with OpenCV and ArUco,
to track fiducials on the network holder.
The accuracy achieved with these methods was never better
than 1cm, and suffered from other problems
such as loss of tracking during motion blur
and occlusion.

First, we tested using the RGB camera in the HoloLens,
for which there is functionality provided to use Vuforia
to track fiducials.
We found this method sometimes exhibited latency
of more than 1 second and sometimes reported
180$^\circ$ errors in the orientation of the fiducials.

\begin{figure}[h!]
\centering
\includegraphics[width=0.5\columnwidth]{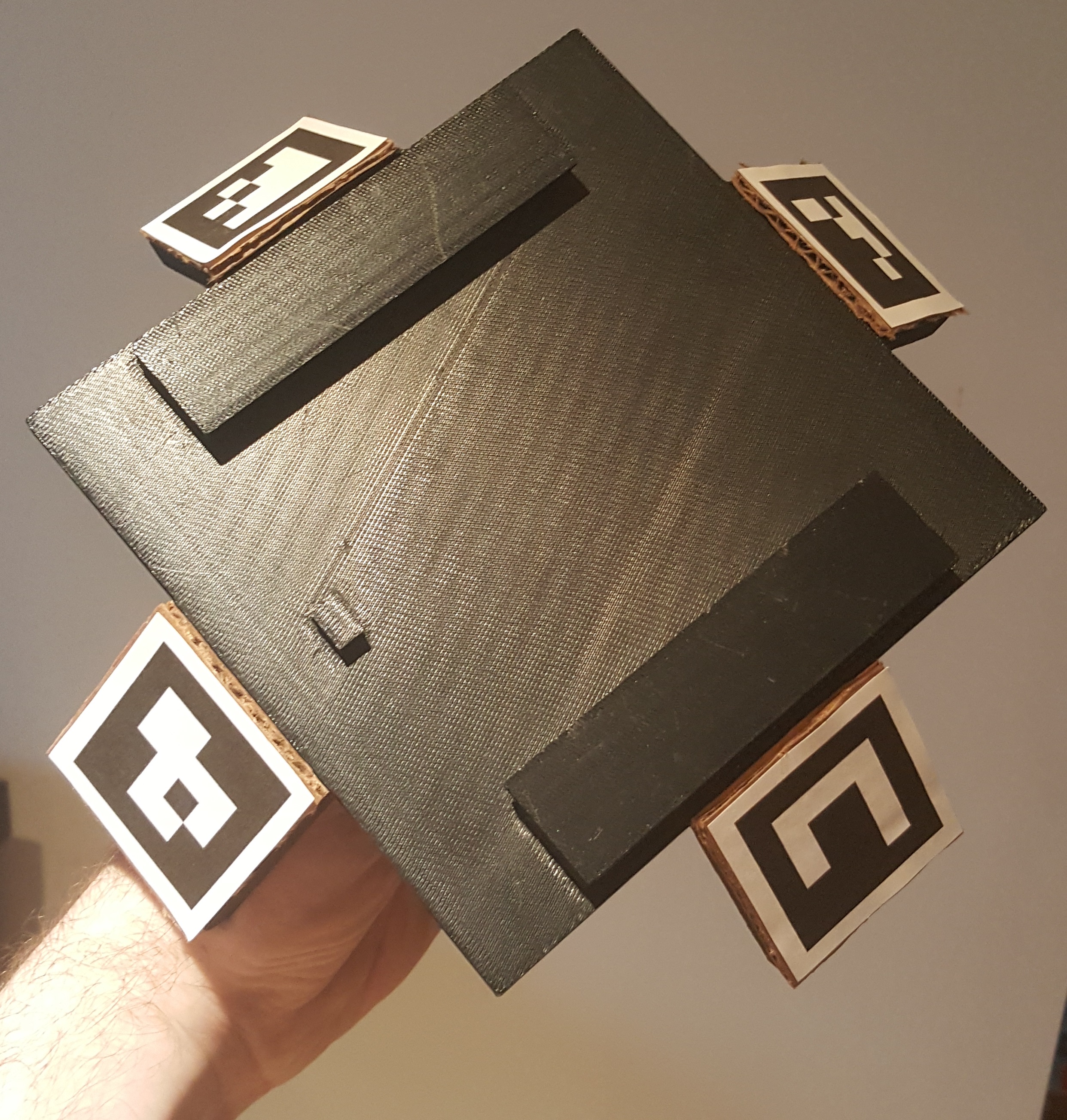}
\caption{Network holder with 4 fiducials.}
\label{fig:netHolderWithFiducials}
\end{figure}

\begin{figure}[h!]
\centering
\includegraphics[width=0.5\columnwidth]{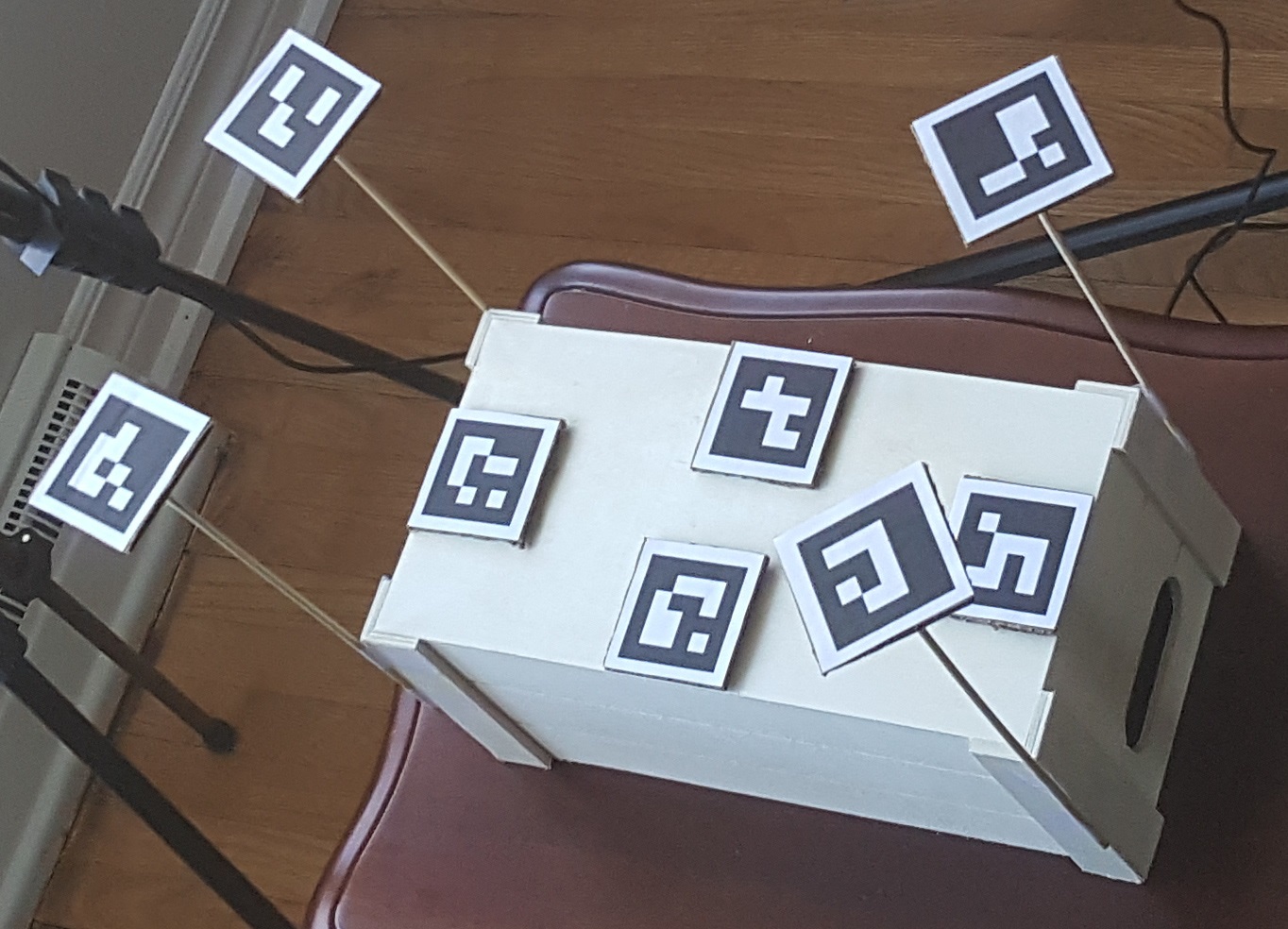}
\caption{Registration rig with 8 fiducials.}
\label{fig:fiducialRig}
\end{figure}

Next, we implemented a tracking system using 3 external Logitech HD (1920$\times$1080) resolution USB cameras
all connected to one PC (the same PC described in the main paper, with a 6-core, 4.7 GHz CPU).
We wrote custom code using OpenCV and ArUco to track fiducials and also to compute 3D-to-3D homographies
to transform from the coordinate system of each camera to the coordinate system of the HoloLens.
We placed 4 fiducials on the network holder
(Figure~\ref{fig:netHolderWithFiducials}),
and also built a physical rig marked with 8 fiducials
(Figure~\ref{fig:fiducialRig})
for registration.
We found that
tracking of fiducials with this method could fail if there were shadows,
which we alleviated using softbox lighting,
and also whenever the fiducials moved too quickly causing a blurry image.
An additional problem was the tradeoff between
spatial and temporal resolution.
Despite using a fast PC with a Vantac ``quad chip'' USB card (with a dedicated chip for each USB port),
we could only read the 3 cameras at $\approx$ 5-10 fps at full HD resolution.
Downgrading the camera resolution to 640$\times$480 allowed reading at 30 fps,
but because each camera has a FOV of 66$^\circ$ horizontally,
the spatial resolution was
(arc length covered by one pixel) = (FOV in radians)$\times$(distance to camera)/(pixels) = ((66$^\circ$/180$^\circ$)$\pi$)$\times$(100cm)/(640 pixels) $\approx$ 0.2 centimeters, assuming the camera is 100cm away from the fiducial.  ArUco's estimation of a fiducial's center seemed very sensitive to image noise, moving 2-3 pixels from frame-to-frame even when the fiducial is not moving.
So this could result in an error of more than 0.5cm,
even assuming no error from the estimated position of
the HoloLens.

Our 3rd attempt to track the network holder was
done with the Polhemus Patriot, as reported
in the main paper.